\documentclass[12pt]{article}
\usepackage{amsmath}
\usepackage{graphicx,psfrag,epsf}
\usepackage{enumerate}
\usepackage{comment}
\usepackage{natbib}
\usepackage{amssymb}
\usepackage[dvipsnames]{xcolor}
\usepackage{bm}
\usepackage{listings}
\usepackage{array}
\usepackage[normalem]{ulem}
\usepackage{multirow}
\usepackage{subcaption}
\usepackage{hyperref}
\usepackage{caption}
\usepackage{booktabs}
\usepackage{longtable}
\usepackage{caption}
\usepackage{setspace}
\usepackage[ruled]{algorithm2e}

\newcommand{\blind}{0}

\newcolumntype{P}[1]{>{\raggedright\arraybackslash}p{#1}}
\newcolumntype{L}[1]{>{\raggedright\let\newline\\\arraybackslash\hspace{0pt}}m{#1}}
\newcolumntype{C}[1]{>{\centering\let\newline\\\arraybackslash\hspace{0pt}}m{#1}}
\newcolumntype{R}[1]{>{\raggedleft\let\newline\\\arraybackslash\hspace{0pt}}m{#1}}

\SetKwInOut{Input}{Input}
\SetKwInOut{Output}{Output\,}
\SetKwProg{stepone}{Step 1: Bin the Data}{}{}

\SetKwProg{steptwo}{Step 2: Estimate Local GLMMs}{}{}
\SetKwProg{stepthree}{Step 3: FPCA on Local Latent Estimates}{}{}

\SetKwProg{stepfour}{Step 4: Re-estimate Subject Specific Scores}{}{}

\begin{document}



\if0\blind
{
  \title{\bf Fast Generalized Functional Principal Components Analysis}
    \author{
    Andrew Leroux\\
    \normalsize{Department of Biostatistics and Informatics, University of Colorado Anschutz Medical Campus} \\ \\
    Ciprian M. Crainiceanu \\
    \normalsize{Department of Biostatistics, Johns Hopkins Bloomberg School of Public Health} \\ \\
    Julia Wrobel \\
    \normalsize{Department of Biostatistics and Informatics, University of Colorado Anschutz Medical Campus}
 }
  \maketitle
} \fi

\if1\blind
{
  \bigskip
  \bigskip
  \bigskip
  \begin{center}
    {\LARGE\bf Fast Generalized Functional Principal Components Analysis}
\end{center}
  \medskip
} \fi

\bigskip
\begin{abstract}
We propose a new fast generalized functional principal components analysis (fast-GFPCA) algorithm for dimension reduction of non-Gaussian functional data. The method consists of: (1) binning the data within the functional domain; (2) fitting local random intercept generalized linear mixed models in every bin to obtain the initial estimates of the person-specific functional linear predictors;  (3) using fast functional principal component analysis to smooth the linear predictors and obtain their eigenfunctions; and (4) estimating the global model conditional on the eigenfunctions of the linear predictors. An extensive simulation study shows that fast-GFPCA performs as well or better than existing state-of-the-art approaches, it is orders of magnitude faster than existing general purpose GFPCA methods, and scales up well with both the number of observed curves and observations per curve. Methods were motivated by and applied to a study of active/inactive physical activity profiles obtained from wearable accelerometers in the NHANES 2011-2014 study. The method can be implemented by any user familiar with mixed model software, though the {\ttfamily R} package {\ttfamily fastGFPCA} is provided for convenience.
\end{abstract}

\noindent%
{\it Keywords:} functional data, FPCA, generalized FPCA
\vfill

\newpage
\doublespacing


\section{Introduction}
\label{sec:introduction}

Functional data analysis (FDA)  \cite{ramsaysilv2005} provides a wide range of analysis techniques for data with complex structures such as time series or images. These data are often high dimensional (many repeated observations per function) and exhibit complex and non-stationary patterns of variation. Functional principal components analysis (FPCA) \cite{10.1080/00031305.1992.10475870,ricesilverman1991,Staniswalis1998}, the analog of principal components analysis (PCA) \cite{Hotelling1933AnalysisOA,jollife1982,10.1080/14786440109462720} for functional data, is a first-line dimension reduction technique for the analysis for such data. Key differences between FPCA and PCA are that functional data: (1)  may be observed with substantial measurement error; (2) is expressed in the same unit of measurement at every point in the domain; and (3) has a functional domain that is ordered and has a natural distance (e.g., time ordering and distance).

There is a rich literature on FPCA focusing on both estimation and inference. Broadly, FPCA methods for Gaussian data involve either smoothing the covariance function  \cite{Staniswalis1998, xiao2016, yao2003} or estimation via model-based approaches with explicit likelihood assumptions \cite{10.1016/j.csda.2008.09.015, james2001}. Extensions to sparse or irregularly observed data \cite{PengPaul2009,yao2005, xiao2018}, multivariate \cite{Happ2015MultivariateFP,10.2307/24310959,lixiaomfaces} and multilevel functional \cite{di2009, 10.1080/10618600.2022.2115500} data exist.  Although the FPCA literature is quite extensive, there are few high quality, open source, software implementations. The covariance smoothing FACE approach of \cite{xiaoface2021,xiao2016}, implemented in the {\ttfamily refund} package \cite{refund} in {\ttfamily R}, is by far the fastest approach for estimating FPCA for regularly observed data. The likelihood-based methods require substantially longer computation times for large data.

Here we focus on extensions of FPCA methods to non-Gaussian outcomes (e.g., binary or count data), which we refer to as Generalized FPCA (GFPCA). More precisely, we focus on methods that decompose the variability of the person-specific latent functional linear predictors along their main directions of variation.  In contrast to the relatively large number of published papers on FPCA, there are far fewer GFPCA papers. A few exceptions are \cite{10.1111/j.1467-9868.2008.00656.x, 10.1007/s10182-009-0113-6, 10.1016/j.csda.2016.07.010,WEISHAMPEL2023107647} for single-level and \cite{10.1080/01621459.2013.826134, doi.org/10.1111/biom.12278} for multilevel GFPCA. Unfortunately, the software implementation problem is even more acute for GFPCA compared to FPCA. Indeed, most published methods either lack accompanying software or are extremely slow for large data.  Moreover, current methods require pre-specifying both the number of principal components and basis functions used to estimate the principal components. Assessing sensitivity to these key input parameters is critical when applying GFPCA to a new dataset.

The ever increasing number of studies that collect non-Gaussian functional data of increasing size and complexity require methods that are fast and scalable. Consider, for example, the minute level physical activity data obtained from wearable accelerometers deployed in large epidemiologic studies, such as the National Health and Nutrition Survey (NHANES) and UK Biobank \cite{10.1371/journal.pone.0169649}. In many applications one is interested in the pattern of being active (coded as $1$) or inactive (coded as $0$) at every minute of the day. Thus, the data at the study participant level is a function observed at $1440$ minutes (number of minutes in a day) where the  value of the function is either active or inactive at every time point. NHANES contains such data for tens of thousands of study participants, while the UK Biobank for close to $100,000$ study participants. Our goal is to provide a scalable GFPCA method that aids in interpretation and analysis of large-scale non-Gaussian functional data from the NHANES accelerometry study by extracting orthogonal directions of variation in the linear predictor space of these functions.

Very recently, \cite{WEISHAMPEL2023107647} proposed an approach which allows for estimation of GFPCA for general exponential family outcomes. Their approach is fast, could be viewed as an alternative, but differs in key ways from fast-GFPCA. Importantly, in our data application, the approach of \cite{WEISHAMPEL2023107647} has the potential to lead to infinite bias in estimated latent functions. We discuss this point further in Section~\ref{subsec:methods_alternatives}, but do not focus on their approach because it was not appropriate for our data application.
Aside from \cite{WEISHAMPEL2023107647}, general purpose software for GFPCA that accommodates multiple types of exponential family data are slow. However, there is also good news about specific types of outcomes. For example,  \cite{10.1111/biom.12963} developed a very fast and efficient binary FPCA procedure that uses an EM algorithm to optimize a variational approximation to the binomial FPCA likelihood. The paper is accompanied by the {\ttfamily registr} package \cite{registr_package, registr2}. To our knowledge this is the only publicly available GFPCA implementation that is fast and viable for large datasets, and thus we compare our fast-GFPCA approach to the {\ttfamily registr::bfpca()} function. In addition, for simulation scenarios with smaller data sets, we compare fast-GFPCA to the two-step GFPCA implementation described in \cite{10.1016/j.csda.2016.07.010}, which is available in the {\ttfamily registr} package as the {\ttfamily registr::gfpca\_twoStep()} function. This two-step approach can be used to model several exponential family outcomes but is prohibitively slow for large datasets. The current work adds substantially to the literature by providing a general approach to GFPCA which is readily generalizable to functional regression models of interest for which there are currently no fast implementations for large scale data.

Specifically, we propose a new method, fast-GFPCA, with an entirely different philosophy and implementation strategy than most other approaches. The advantages of fast-GFPCA are that: (1) it can be used for any type of non-Gaussian outcome, not just binary; (2) it is orders of magnitude faster than most other all-purpose GFPCA approaches; (3) the method readily handles missing data; (4) it can easily be extended to account for covariates; and (5) it can be generalized to multilevel, longitudinal or structured functional data. We will briefly discuss these extensions, but leave the details for future work.

The remainder of this manuscript is organized as follows. In Section~\ref{sec:methods} we present the fast GFPCA (fast-GFPCA) approach. Next, in Section~\ref{sec:application}, we apply the fast-GFPCA to active/inactive profiles obtained from wearable accelerometers using data from the National Health and Nutrition Survey (NHANES) 2011-2014 waves. We then illustrate the utility of the fast-GFPCA approach in a simulation study in Section~\ref{sec:simulation}. We conclude with a discussion in Section~\ref{sec:discussion}.

\section{Methods}
\label{sec:methods}
The observed data structure is of the type $\{s_j,Z_i(s_j)\}$, where $Z_i(s_j)$ is a non-Gaussian functional observation for subject $i \in 1,\ldots, N$
at the point $s_j\in S$ for $j \in 1,\ldots, J$. We assume that these $\{s_j,Z_i(s_j)\}$ pairs are discrete realizations from a continuous process $\{Z_i(s):s\in S\}$ such that: (1) $g[E\{Z_i(s)\}] = \beta(s) + b_i(s)$, where $g(\cdot)$ is an appropriate link function, $\beta(s)$ is the population mean function in the linear predictor space, and $b_i(s)$ is the individual deviation from the population mean in the linear predictor space; (2) $b_i(s) \sim \text{GP}(0, K_b)$ is a zero mean Gaussian process with covariance operator $K_b$. Our goals are to decompose the variability of the latent process $b_i(s)$ along its main directions of variation (obtain the FPCA decomposition) and estimate $b_i(s)$ conditional on these directions of variation. Even though $b_i(s)$ are not directly observed, we can use the Karhunen-Lo{\`e}ve (KL) expansion $b_i(s) = \sum_{k=1}^\infty \xi_{ik}\phi_k(s)$ where $\phi_k:S\rightarrow \mathbb{R}$ are orthonormal eigenfunctions such that $\int_S\phi_k^2(s)ds=\lambda_k$, $\lambda_1\geq \lambda_2\geq \ldots$ are the eigenvalues, and  $\xi_{ik}{\sim} N(0, \lambda_k)$ are mutually independent subject-specific scores over study participants, $i$, and eigenvalues, $k$. Together this leads to the GFPCA model,
\begin{align}
                g\left(E[Z_i(s)]\right)  &= \eta_i(s) = \beta_0(s) + \sum_{k=1}^{\infty} \xi_{ik}\phi_k(s)\;.
            \label{eq:gfpca_model}
\end{align}
\noindent This is very similar to the classical FPCA model with the exception that the noise is not necessarily Gaussian.

\subsection{Fast GFPCA}
\label{subsec:methods_fastGFPCA}
We propose the following fast-GFPCA algorithm to conduct FPCA on latent processes when observed data are non-Gaussian: (1) bin the data along the observed functional domain $\mathcal{S} = 1,\ldots,J$ into $L$ bins which may be overlapping; (2) estimate separate local GLMMs with a random intercept in each bin to obtain subject-specific estimates on the linear predictor scale; (3) estimate FPCA on the subject-specific estimates obtained from step (2); (4) re-estimate GFPCA using the estimated eigenfunctions obtained from step (3) at the resolution of the original data. 
Each of these steps is straightforward, though we provide the {\ttfamily R} package {\ttfamily fastGFPCA} for convenience, described in Section~\ref{subsec:methods_code}. Details on each step are provided in Section~\ref{subsec:methods_estimation} below.

\subsection{Estimation Algorithm}
\label{subsec:methods_estimation}

We now provide more details on each of the four steps of the fast-GFPCA algorithm, while the algorithmic presentation is provided in Section~\ref{sec_supp:algorithm} of the supplementary material.

\paragraph{Step 1:} Choose how to bin the data. The choice of the number of bins ($L$) and the bin widths ($w_l$) will be informed by identifiability and assumed complexity of the underlying latent process.  Specifically, suppose we choose $L$ bins, where $m_l$ is the midpoint of the $l=1,\ldots,L$ bin. We use symmetric bins, except on the boundary of the domain $S$. For data observed on a regularly spaced grid over the domain, the $l^{\text{th}}$ bin contains the data at domain values $\mathcal{S}_l = \{s_{m_l-w_l/2},\ldots, s_{m_l}, \ldots, s_{m_l+w_l/2}\}$, resulting in $w_l + 1$ points. The binned data are then $[\{Z_i(s_j), l\}, 1 \leq i \leq N, j \in \mathcal{S}_l, 1 \leq l \leq L ]$. 

If the data are cyclic, as in our data application, the bins may cross the boundary. For example, with minute level accelerometry data, if we let $m_1 = 1$ (activity 00:00-00:01) and $w_l = 6$, then $\mathcal{S}_1 = \{1438, 1439, 1440, 1, 2, 3, 4\}$ (activity 23:57-00:05). When the data are non-cyclic we recommend constructing bins as $\mathcal{S}_l = \{\text{min}(s_{m_l-w_l/2}, s_1),\ldots, s_{m_l}, \ldots,$ $\text{max}(s_{m_l+w_l/2}, s_J)\}$, resulting in bins with as few as $w_l/2 + 1$ data points. 

\paragraph{Step 2:} Fit a local GLMM in every bin. Specifically, in each bin $l=1,\ldots,L$ we estimate separate models of the form $g[E\{Z_i(s_j)\}] = \beta_0(s_{m_l}) + b_i(s_{m_l}) = \eta_i(s_{m_l})$ for $s_j \in \mathcal{S}_l$. Here $\beta_{0}(s_{m_l})$ is a fixed effect mean and $b_i(s_{m_l})$ is a random intercept evaluated at the center of the bin, $s_{m_l}$. From these models we obtain estimates of the global mean, $\widehat{\beta}_0(s_{m_l})$, and predictions of the subject-specific random effects $\widehat{b}_i(s_{m_l})$. These predictions are {\it not} on the original grid the functions were observed on, but rather the midpoints $\{s_{m_1},\ldots,s_{m_L}\} \subset \mathcal{S}$. Importantly, this model is misspecified because it assumes a constant effect in each bin while, in reality, both $\beta_0(\cdot)$ and $b_i(\cdot)$ vary smoothly over $\mathcal{S}$. This can lead to biased estimates for $\beta_0(s_{m_l})$, $b_i(s_{m_l})$,
 which has the potential to induce bias in our estimator for $K_b$. This bias, under reasonable choice of bin width, is largely absorbed in the eigenvalues of the estimated covariance operator, with eigenfunctions being well estimated. As the method hinges on estimating the eigenfunctions well in step 3, but not necessarily the eigenvalues, this is not a problem for the method. Below, we deviate briefly from the description of the fast-GFPCA algorithm to discuss the issue of bias in more detail as we believe such discussion provides critical insights into why the proposed method works in practice and is reasonable.

\paragraph{Discussion of Binning Induced Bias in Estimation of $\mathbf{K_b}$} We argue that the binning procedure induces bias in the estimated latent functions evaluated at the midpoints of each bin which in turn induces bias in the estimated covariance function $K_b$. The point is most readily shown when we assume that the distribution of observed data are continuous uniform along the domain. For ease of presentation, assume that the domain is $S = [0,1]$. It follows that the distribution of points follows density $f_s(s) = 1$, $s \in [0,1]$. Further, conditional on $s \in S_l$ ($s$ is in the $l^{\text{th}}$ bin), the density of points is iid uniform with conditional density $f_{s|l}(s) = |S_l|^{-1}$ (inverse of the interval width).

Let the superscript $^{\text{bin}}$ notation denote the estimand under the misspecified model to differentiate between the ``true" latent process. The misspecified model in Step 2 estimates $\eta_i^{\text{bin}}(s_{m_l}) = E[\beta_0(s) + b_i(s)|\{\xi_{ik}: 1 \leq k \leq K\}, s \in \mathcal{S}_l] = E[\beta_0(s)|s \in \mathcal{S}_l] + E[b_i(s)|\{\xi_{ik}: 1 \leq k \leq K\}, s \in \mathcal{S}_l]$, with bias which can be split into population $\text{Bias}\left[\beta_0^{\text{bin}}(s_{m_l})\right] = \beta_0(s_{m_l}) - E[\beta_0(s)|s \in \mathcal{S}_l] $ and subject-specific $\text{Bias}\left[b_i^{\text{bin}}(s_{m_l})\right] = b_i(s_{m_l}) - E[b_i(s)|\xi_{ik}, s \in \mathcal{S}_l]$ components. The key to fast-GFPCA is the ability to obtain reasonable estimates for the covariance operator $K_b = \text{Var}\left[b_i(s)\right]$ in Step 3. The additive bias in $\beta_0^{\text{bin}}(s_{m_l})$ does not affect the estimator of the covariance operator in Step 3 as the data (i.e. $\text{Cov}(a + X, Y) = \text{Cov}(X,Y)$), so we focus on the effect of the subject specific bias, given by    
\begin{eqnarray*}
\text{Bias}\left[b_i^{\text{bin}}(s_{m_l})\right] &=& b_i(s_{m_l}) - E[b_i(s)|\xi_{ik}, s \in \mathcal{S}_l]\\
&=& \sum_{k = 1}^K \phi_k(s_{m_l})\xi_{ik} - E[\sum_{k=1}^K \phi_k(s)\xi_{ik}| \xi_{ik}, s \in \mathcal{S}_l]\\
&=& \sum_{k = 1}^K \phi_k(s_{m_l})\xi_{ik} - \sum_{k=1}^K \xi_{ik} \int_{s \in S_l} \phi_k(s) f_{s|l}(s) ds\\
&=& \sum_{k = 1}^K \phi_k(s_{m_l})\xi_{ik} - \sum_{k=1}^K \xi_{ik} |S_l|^{-1}\int_{s \in \mathcal{S}_l} \phi_k(s)ds
\end{eqnarray*}
with $|S_l|^{-1}$ operating as a normalizing constant so that $\lim_{w_l \to 0}|S_l|^{-1}\int_{s \in \mathcal{S}_l} \phi_k(s)ds = \phi_k(s_{m_l})$. That is, as bin width tends to $0$ we recover the eigenfunction evaluated at the midpoint of the current bin. Then 
\begin{align*}
    \text{Cov}\left[b_i^{\text{bin}}(s_{m_u}), b_i^{\text{bin}}(s_{m_v})\right] &= \sum_{k=1}^K \lambda_k \left[|S_u|^{-1}\int_{\mathcal{S}_u}\phi_k(s) ds\right]\left[|S_v|^{-1}\int_{\mathcal{S}_v}\phi_k(s)ds \right] \\
    &\approx \sum_{k=1}^K \{\lambda_k |S_u|^{-1}|S_v|^{-1}\} \phi_k(s_{m_u}) \phi_k (s_{m_v})\;.
\end{align*}
\noindent
The last approximation is related to the mean value theorem, which states that $\frac{1}{|S_u|}\int_{S_u}\phi_k(s)ds$ $\approx \phi_k(s_{m_u})$ and $\frac{1}{|S_v|}\int_{S_v}\phi_k(s)ds\approx \phi_k(s_{m_v})$ so long as $|S_u|$, $|S_v|$ are not large relative to the curvature in $\phi_k$. If the function $\phi_k(\cdot)$ is constant or linear in any of these intervals, the approximation is actually an equality. This equation shows that by diagonalizing $\text{Cov}\left[b_i^{\text{bin}}(s_{m_u}), b_i^{\text{bin}}(s_{m_v})\right]$ we obtain close approximations of the eigenvectors $\phi_k(\cdot)$, but not of the eigenvalues $\lambda_k$. Indeed, the eigenvalues are re-scaled by $|S_u||S_v|$, which is the square of the length of the approximating interval when using constant bin width. This also explains why all the bias in the covariance estimation is absorbed by the eigenvalues and passed on, implicitly, to the scores. Therefore, if we diagonalize the linear predictors obtained in this step we obtain unbiased eigenfunctions but biased covariance, eigenvectors, scores, and subject-specific trajectories.

\paragraph{Step 3:} Use the predicted responses on the linear predictor scale $[\{\widehat{\eta}_i(s_{j}), l\}, 1 \leq i \leq N, j \in \mathcal{S}_l]$ to estimate $\widehat{K}_b(u,v)=\text{Cov}\{\widehat{\eta}_i(u), \widehat{\eta}_i(v)\}$ for $u,v \in \{s_{w_1},\ldots, s_{w_L}\}$ via the fast covariance estimation (FACE) method implemented in the {\ttfamily refund::fpca.face()} function. Obtain the estimated eigenfunctions, $[\{\hat{\phi}_k(s_{w_l})\}, 1 \leq l \leq L, 1 \leq k \leq K]$ of the covariance operator $\widehat{K}_b$ where the number of eigenfunctions, $K$, is selected using, for example, the percent variance explained (e.g., 95\%, 99\%,).

\paragraph{Step 4:} Estimating the GFPCA model conditional on the basis functions obtained in Step 3. Because the basis functions are estimated on a different grid from the one of the observed data, we first project each eigenfunction on the rich B-spline basis used in the FACE component of the algorithm in Step 3. This provides an estimate of the eigenfunctions at every point where data was originally sampled. After these projections, fast-GFPCA becomes the following generalized linear mixed model (GLMM) 
\begin{equation}
            g(E[Z_i(s)]|\{\hat{\phi}_k(s): 1 \leq s \leq J, 1 \leq k \leq K \})  = \beta_0(s) + \sum_{k=1}^K \xi_{ik}\hat{\phi}_k(s)\;,
            \label{eq:full_conditional}
\end{equation}
where $\xi_{ik} \sim N(0,\sigma_k^2)$ are mutually independent and $\beta_0(s)$ is an unspecified smooth function. That is, given the estimates of $\hat{\phi}_k(\cdot)$, this is a GLMM with $K$ uncorrelated random slopes. Using a principal components decomposition with uncorrelated random slopes simplifies the random effects covariance structure such that only $K$ variance parameters need to be estimated, which contributes to computational efficiency.

If we assume a parametric form for $\beta_0(s)$, any generalized linear mixed model software can be used. For the case when $\beta_0(s)$ is modeled nonparametrically, we estimate Model \ref{eq:full_conditional} using the {\ttfamily mgcv::bam()}  function \cite{10.1080/01621459.2016.1195744} with fast REML smoothing parameter selection and the argument {\ttfamily discrete=TRUE} \cite{10.1007/s11222-019-09864-2}. This approach is highly computation efficient; see example code in Section~\ref{subsec:methods_code}. Alternative software for estimating additive generalized linear mixed models are available and may be faster and more memory efficient in certain situations; specifically, the {\ttfamily mgcv::gamm()} \cite{10.1111/j.1467-9868.2010.00749.x} and {\ttfamily gamm4::gamm4()} \cite{10.1007/s11222-012-9314-z} functions, which provide interfaces between the {\ttfamily mgcv} and {\ttfamily nlme} \cite{nlme} and {\ttfamily lme4} \cite{lme4} packages, respectively.

\subsection{Fast GFPCA: Example Code}
\label{subsec:methods_code}

A key appeal of fast-GFPCA is that it may be implemented by anyone familiar with mixed model software. The code below illustrates how fast-GFPCA can be implemented on a subset of the binary NHANES active/inactive profiles data used in the application described in Section \ref{sec:application}. For illustration we use overlapping windows and bin width $w_l = 10$ for $l = 1,\ldots, 1440$ (both $w_l$ and $l$ are expressed in minutes). The general organization of the code follows the four steps of the fast-GFPCA algorithm. This code can also be run using a one-line implementation with the accompanying {\ttfamily fastGFPCA} package. 

\small
\singlespacing
\begin{verbatim}
library("refund"); library("tidyverse"); library("lme4")
df      <- read.rds("NHANES_example.rds") # read in the data
J       <- length(unique(df$index))       # dimension 

## Step 1: Binning decisions
bin_len <- 10                             # bin width (w)
s_m     <- 1:J                            # bin midpoints s_{m_l}

## Step 2: Fit local binned GLMMs
fit_ls  <- vector(mode="list", length=J)  # empty list to store results
for(j in s_m){                            # loop over bins
        # get indices associated with current bin S_l
        # Note: this data is cyclic, so we look across the domain
        #       using the modulo (%%) function
        sind_j <- (j-bin_len_w/2):(j+bin_len_w/2) %% 1440    
        sind_j[sind_j == 0] <- 1440
        # subset to the current indices
        df_j <- df %>% 
                filter(index %in% sind[sind_j])
        # fit the local, binned GLMM:
        #   g(E[Z_i(s)]|s_j \in S_l) = \beta_0(s_{m_l}) + b_i(s_{m_l})
        fit_j <- glmer(value ~ 1 + (1|id), data=df_j, family=binomial)
        # store results (id, \hat{eta}_i(s_{m_l}), s_{m_l})
        fit_ls[[j]] <- data.frame("id" = 1:N, 
                                  "eta_i" = coef(fit_j)$id[[1]],
                                  "s_m" = j)
}
## bind elements of the list row-wise
fit_df <- bind_rows(fit_ls)

## Step 3: FPCA on the binned latent estimates
fpca_latent <- fpca.face(matrix(fit_df$eta_i, N, J, byrow=FALSE), 
                         pve=0.95, argvals=sind, knots=20,lower=0)
                         
## Step 4: Re-fit the GLMM on the full data
##         Note that here we do not need to interpolate the eigenfunctions
##         as the bin midpoints contained the original observations points
# data frame of eigenfunctions (take the first four)
phi_mat <- data.frame("index" = s_m, fpca_latent$efunctions[,1:4])
colnames(phi_mat)[2:5] <- paste0("Phi", 1:4)
# merge the data 
df <- df %>% left_join(phi_mat, by="index") %>% mutate(id_fac=factor(id))
# re-fit the model using mgcv::bam()
fit <- bam(value ~ s(index,bs="cc",k=20) + 
                   s(id_fac,by=Phi1,bs="re") + s(id_fac,by=Phi2,bs="re") + 
                   s(id_fac,by=Phi3,bs="re") + s(id_fac,by=Phi4,bs="re"), 
           data=df, family=binomial, discrete=TRUE, method="fREML")
\end{verbatim}
\normalsize
\doublespacing

For presentation simplicity, the code assumes that data are ordered by subject and then by the domain of the function. If the data are not sorted like this, slight modifications are needed in Steps 3 and 4 or data can be re-ordered before the code is run. Moreover, the subject identifier needs to be a factor variable to fit the correct model, which is why the variable {\ttfamily id\_fac} was created in Step 4. Users unfamiliar with the {\ttfamily mgcv} package may be confused by the syntax in the call to {\ttfamily mgcv::bam()} in Step 4. The expression {\ttfamily s(id\_fac, by=Phi1, bs="re")} constructs independent normally distributed random slopes. The same random effects specification in {\ttfamily lme4} would be

\singlespacing
\begin{verbatim}
fit <- glmer(value~1 + (0+Phi1|id) + (0+Phi2|id) + (0+Phi3|id)+(0+Phi4|id),
             data=df, family=binomial),
\end{verbatim}
\doublespacing
however, {\ttfamily lme4::glmer()} cannot model $\beta_0(t)$ nonparametrically and a parametric form for $\beta_0(s)$ would need to be specified. 
We hope that seeing the complete code will: (1) show how easy it is to implement the proposed methods; (2) provide a modular and modifiable platform that can be used in similar situations which require specific adjustments; and (3) support the philosophy that analytic methods are not really methods without supporting software. 

For  user convenience we have developed {\ttfamily fastGFPCA}, an {\ttfamily R} package available on GitHub at \url{https://github.com/julia-wrobel/fastGFPCA}, which wraps the code for implementing fast-GFPCA. The detailed description of the package functionality, including examples for both binomial and Poisson distributed functional data, is contained in the \texttt{fastGFPCA} vignette associated with the package. Briefly, the primary function in the package is {\ttfamily fast\_gfpca()}. Function arguments {\ttfamily overlap} and {\ttfamily binwidth} allow the user to select a bin width and choose whether or not to construct overlapping windows for their data. The argument {\ttfamily family} is fed to the functions {\ttfamily lme4::glmer()} and {\ttfamily mgcv::bam()} and is used to specify an exponential family distribution and link function using the syntax, for example, {\ttfamily family = binomial(link = "logit")}. The {\ttfamily fast\_gfpca()} function returns an object of class {\ttfamily "fpca"} and results can be easily visualized using the {\ttfamily refund.shiny} package \cite{refundshiny}.

\subsection{Practical Considerations}
\label{subsec:methods_considerations}

Several practical considerations must be taken into account when applying the fast-GFPCA method. We discuss these considerations in detail below.

\paragraph{The Need for Step 4.} The need for Step 4 stems from the fact that Step 3 produces unbiased estimators of the eigenfunctions but biased estimators of the subject-specific random effects. This is driven primarily by bias in the estimated eigenvalues of the covariance operator, which are then passed onto the scores. This effect can be seen in Figure~\ref{fig:need_for_step_4}, which plots the estimated scores from step 3 and step 4 versus the true scores (Figure~\ref{fig:need_for_step_4}(A)), the estimated scores from step 3 versus those from step 4 (Figure~\ref{fig:need_for_step_4}(B)), and the estimated curves on the log odds scale using step 3 versus step 4. The data are generated as binary functional data according to our simulation study (See Section~\ref{sec:simulation} for details), and fast-GFPCA estimates the true eigenfunctions well in this data. From Figure~\ref{fig:need_for_step_4}(A) and Figure~\ref{fig:need_for_step_4}(B), we see that not only are the scores obtained from step 3 biased (far away from the identity line, Figure~\ref{fig:need_for_step_4}(A)), but that the scores in step 4 are effectively a linear re-scaling of those from step 3 (near perfect correlation, but points away from the identity line, Figure~\ref{fig:need_for_step_4}(B)). This bias results in estimated log-odds which are substantially less correlated with the true latent functions (Figure~\ref{fig:need_for_step_4}(C)). Note that the slopes in Figure~\ref{fig:need_for_step_4}(B) are nearly identical which matches the result shown in Section~\ref{sec:methods} Step 2 (discussion on binning bias) that the eigenvalues are biased by a constant multiplicative factor (and thus the scores are as well).

\paragraph{Identifiability.} Local GLMMs may be non-identifiable or model fitting may not converge. An example is when data are binary and all (or nearly all) observations are either $0$ or $1$ for every study participant in a particular bin. In this case the local model is non-identifiable. Potential solutions include choosing a wider bin width, imposing a lower or upper bound on the linear predictor scale, or modifying the data. For example,  \cite{10.2307/2685779} showed that adding two successes and two failures improves the performance of confidence intervals when estimating a probability from binary data. The idea can be used in our context by adding two successes and two failures in every bin. For our application and simulations, increasing the bin size was enough to ensure excellent performance of the methods.

\paragraph{Bin width.} As mentioned in Section~\ref{sec:methods}, the choice of bin width is an important tuning parameter in the fast GFPCA algorithm. Our simulation study presented in Section~\ref{sec:simulation} illustrates this point. There is a need to balance choosing a bin size that is small enough to estimate the curvature of the latent process but not too small to make GLMM fitting unstable.  A possible approach is to consider bins of increased sizes to the level where GLMMs can fit the data. The bin sizes can be increased and the stability of estimators compared. An alternative would be to conduct smoothing at the study specific level and inspect the plots to explore the complexity of the underlying functions.

\paragraph{Non-overlapping versus overlapping windows.} The choice of whether to use non-overlapping versus overlapping windows will vary by application. Non-overlapping windows reduces the number of local GLMMs that need to be estimated, though computational gains are minor. In contrast, overlapping windows allows estimation at a finer grid of points $s_{m_1},\ldots,s_{m_L}$, but may induce spurious correlation is the estimated latent processes. The effects of these auto correlations, if any, are not currently known in applications, though we illustrate that the method can result in under-smoothing in our data application. A possible technical solution could be to adapt the smoothing selection criteria of FACE to allow for auto correlated data, though this exceeds the scope of our current work.

\subsection{Alternative Approaches}
\label{subsec:methods_alternatives}

There are at least two alternatives to the proposed approach based on first smoothing individual curves. First, one may consider first smoothing the binary functional data on the response scale and ignoring the binary nature of the data. These smoothed data may then be transformed using a link function (e.g., logit for binary data), and then apply fPCA to the estimates on the latent scale. Alternatively, smoothing may be done by fitting generalized additive models $g(E[Y_i(s)]) = \beta_{0i} + f_i(s)$ separately for each individual, allowing the linear predictor to vary smoothly over the domain where $f_i(s)$ is modelled using a rich basis. fPCA may then be applied to the resulting estimates on the latent scale. The latter approach was recently proposed by \cite{WEISHAMPEL2023107647}. Both of these approaches are intuitive and may be faster in certain scenarios. While they could be used as quick exploratory tools, we will point out some of their hidden drawbacks. 

Consider the first approach. If smoothing is done on the response scale and we denote the smoothed response as $\tilde{Y}_i(s)$, a fundamental requirement is for $g\{\tilde{Y}_i(s)\}$ to be finite and defined. For example, for binary data one would need $\tilde{Y}_i(s) \in (0,1)$ and preferably farther away from the boundaries to ensure that $\text{logit}\{\text{Pr}(\tilde{Y}_i(s)=1)\}]$ is defined and well behaved. In areas of where there are many zeros (e.g., during the night for physical activity) and ones (during the morning for physical activity), this smooth estimators will be $0$ or $1$, respectively. Thus, the smooth estimates would need to be bounded away from $0$ and $1$ using artificial tuning parameters to inverse transform them. This makes the approach dependent on individual choices and may introduce infinite bias due to arbitrary choices of bounding parameters. 

Regarding the second approach, the individual models may not be identifiable. Indeed, consider binary functional data where large regions of the domain are all $0$ or $1$. In these regions, a generalized additive model is identified only due to the smoothness assumption on the coefficient function $f_i(s)$, which may result in the divergence of the estimated log odds. Moreover, all smoothing assumptions are done at the individual level, without taking into account the information from the other subjects. In our data application,  many participants are completely inactive during the very early morning hours (2AM-5AM). In contrast, the fast-GFPCA approach borrows strength from the other study participants to provide reasonable estimates at the subject level in areas where there is very little information for many, but not all, study participants.   

In addition to the issues mentioned above, both approaches involve tuning parameters related to the estimation method for individual model fits. For example, in the context of penalized splines, one must choose the number of splines, the basis, and a method for smoothing parameter selection. The latter point seems especially important given that the generative model implies the amount of ``wiggliness" of each $i = 1,\ldots, N$ latent function is the same. Fitting separate models does not enforce this constraint and may lead to some functions being estimated to be perfectly smooth, while others are quite wiggly. Moreover, while fast-GFPCA can deal with substantial missing areas in a data set, the two alternative methods described here cannot, especially when combined with excess zeros or ones.

However, the most important drawback of both these methods was that we were not able to successfully use them in our NHANES application; see an extensive discussion of these points in Section~\ref{sec:supp_Weishampel_NHANES} of the supplemental material. Moreover, fast-GFPCA can easily be extended to covariate adjusted and multilevel/structured generalized functional data.  It is not immediately clear that either of the two approaches referenced above are readily extended to these scenarios.


\section{Application}
\label{sec:application}

\subsection{NHANES Accelerometry Data}
\label{subsec:NHANES_data_description}

The National Health and Nutrition Examination Survey (NHANES) is a large, ongoing study which provides a nationally representative sample of the non-institutionalized US population. NHANES is conducted by the Centers for Disease Control (CDC) and collects data in two-year waves with the goal of providing information on the health and nutrition of the US population. Wearable accelerometers were deployed in the 2003-2004, 2005-2006, 2011-2012, and 2013-2014 waves of NHANES. 

The 2003-2006 accelerometry component of the NHANES study involved participants wearing a waist-worn accelerometer during waking hours. A guide to analyzing these data is provided in \cite{10.1007/s12561-018-09229-9} and an {\ttfamily R} data package, {\ttfamily rnhanesdata} \cite{rnhanesdata} is publicly available on Github at \url{https://github.com/andrew-leroux/rnhanesdata}. The 2011-2014 accelerometer data, released in 2021, are provided at multiple resolutions: subject, minute, and sub-second level. The subject and minute level data summarize individuals' acceleration patterns based on the new Monitor Independent Movement Summary (MIMS) unit \cite{10.1123/jmpb.2018-0068}. Here, we use the 2011-2014 minute level MIMS data to construct active/inactive profiles for participants.

\sloppy To obtain binary active/inactive profiles, we first threshold participants' daily MIMS data as $Y^B_{ih}(s) = 1\{Y_{ih}(s) \geq 10.558\}$, where $Y_{ih}(s)$ corresponds to the $i^{\text{th}}$ individual's MIMS unit on day $h$ at minute $s$. We then define their active/inactive profile as $Z_i(s) = \text{median}\{Y^B_{ih}(s): h = 1,\ldots, H_i \}$. For example, if $H_i=7$, $Z_i(s)$ is $0$ if study participant $i$ was inactive at time $s$ for at least $4$ days and $1$ otherwise. When the number of good days is even, say $\frac{H_i}{2}$, the median is defined as the $\frac{H_i}{2}+1^{\text{th}}$ largest observation. The threshold for active/inactive on the MIMS unit scale is chosen to be $10.558$, as suggested in \cite{doi/10.2196/38077}, though our methodology would apply similarly to other thresholds. The analytic sample contains data from $N=4286$ participants with 1440 observations per person (minutes in a day). 

Although NHANES is a nationally representative sample, obtaining nationally representative estimates for population quantities and model parameters requires the use of survey weights and survey design \cite{KornandGraubard2011,Lumley2004, skinner2017}. The intersection of survey statistics and functional data analysis is a relatively new area of research  \cite{cardot2013comparison,cardot2013convergence,  cardot2014, 10.1111/biom.13696} and is beyond the scope of the current work. Thus we do not account for the NHANES survey design in our data application, though it is an important direction for future methodological development.

\subsection{Comparison methods and criteria}
\label{subsec:application_approaches}

We apply fast-GFPCA (labeled {\ttfamily fastGFPCA}) to the NHANES data using bin widths ($w_l$) of $6$, $10$, and $30$ minutes, and both overlapping and non-overlapping intervals. We compared methods to the fast binary variational FPCA (labeled {\ttfamily vbFPCA}) implementation in the {\ttfamily registr::bfpca()} function \cite{registr_package,10.1111/biom.12963}. We also consider a modified version of fast-GFPCA (labeled {\ttfamily modified fastGFPCA}) which further speeds up {\ttfamily fastGFPCA} in Step 4 by fitting the model~\eqref{eq:full_conditional} to four sub-samples of the data. The {\ttfamily modified fastGFPCA} approach is described in more detail in Section~\ref{sec_supp:speed_up} of the supplemental material. The approach described in \cite{10.1016/j.csda.2016.07.010} was not computationally feasible for the NHANES data. To facilitate comparisons across models, we fix the number of principal components across all methods to $K = 4$.  

For each approach we compare model parameters and predictive performance. For model parameters we compare the first four estimated eigenfunctions and the population mean function. For predictive performance, we compare the estimated in-sample log-loss associated with each model fit and the estimated area under the receiver operating curve (AUC). Finally, we compare computation times across methods. Though Step 2 of fast-GFPCA could easily be parallelized, computation times are reported for serialized fitting of the models to provide an upper bound for this step. 

Substantial differences were identified in terms of estimated population means and eigenfunctions using the vbFPCA approach and the fast-GFPCA approach (see commentary in Section~\ref{subsec:application_results}). To investigate these differences a brief simulation study based on the NHANES data was conducted, described and summarized in Section~\ref{sec_supp:data_driven_sim} of the supplemental material. 

\subsection{Results}
\label{subsec:application_results}

\subsubsection{Data Application}

\paragraph{Model Parameters.} Figure~\ref{fig:NHANES_results_estimated_parameters} displays the estimated population means $\hat{\beta}_0(s)$ (Figure~\ref{fig:NHANES_results_estimated_parameters}A), and first four eigenfunctions $\hat{\boldsymbol{\phi}}(s)$ (Figure~\ref{fig:NHANES_results_estimated_parameters}B) of the latent process. Each column of Figure~\ref{fig:NHANES_results_estimated_parameters} corresponds to a different model fit where color indicates approach (vbFPCA in red, columns 1-2, and fast-GFPCA in blue, columns 3-8). Modified fast-GFPCA estimates a population mean function for each of four randomly selected sub-samples, which are then averaged to produce the overall $\hat{\beta}_0(s)$. These sub-sample estimates of $\beta_0(s)$ are shown as dashed lines in Figure~\ref{fig:NHANES_results_estimated_parameters}A, columns 3-8. As the modified fast-GFPCA uses the population level $\hat{\boldsymbol{\phi}}(s)$, there are no corresponding dashed plots for the eigenfunctions. 

\begin{figure}[!ht]
    \centering
    \includegraphics[width=\textwidth]{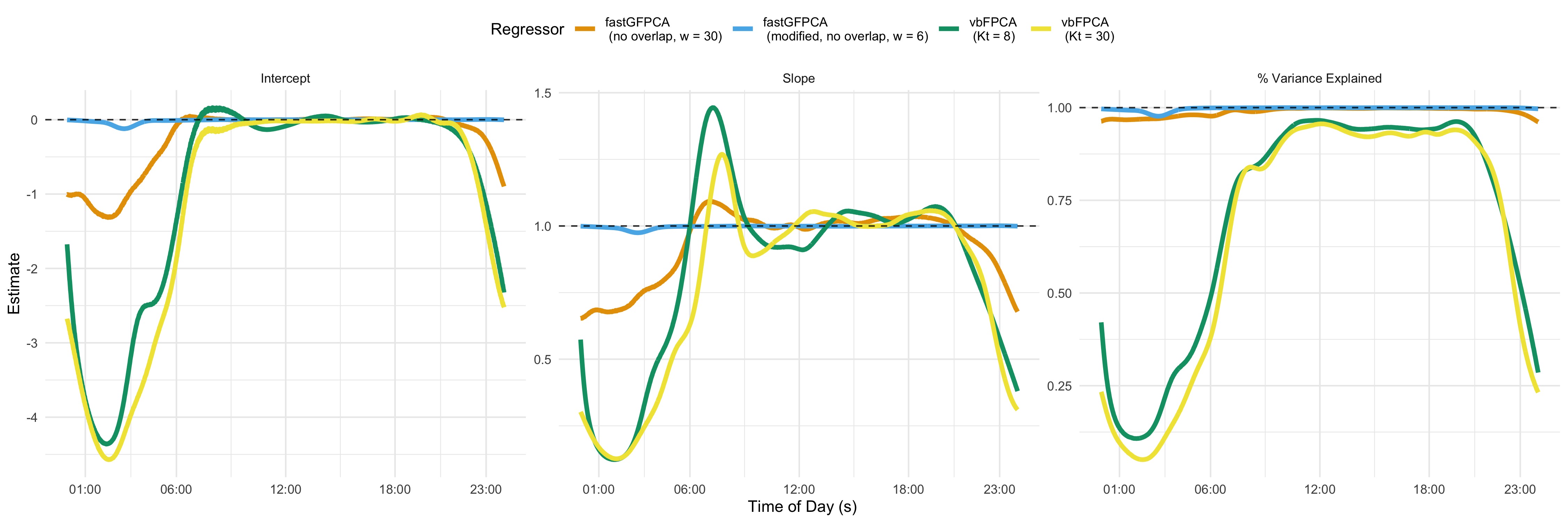}
    \caption{
    {Results from time-specific regression of the estimated log odds of being active obtained from fast-GFPCA using $w_l =6$ with non-overlapping windows and: (1) fast-GFPCA using $w_l =30$ with non-overlapping windows (red lines); (2) vbFPCA with Kt = 8 (green lines); and (3) vbFPCA with Kt = 30 (blue lines). Regressions of the form $E[Y] = \mathcal{B}_0 + \mathcal{B}_1 X$ were fit separately for each minute, with the resulting estimates $\hat{\mathcal{B}}_0$, $\hat{\mathcal{B}}_1$, and the percent variance explained ($R^2$) plotted separately in each panel (left to right). Deviations from the black dashed line in each panel denote a reduced rate of agreement between the regressor and the results from fast-GFPCA using $w_l =6$ with non-overlapping windows.}
    \label{fig:NHANES_corr_linear_predictor}
    }
\end{figure}

The population mean is fairly stable across sub-samples for the modified fast-GFPCA approach (Figure~\ref{fig:NHANES_results_estimated_parameters}A, columns 3-8). We also find  excellent agreement between the estimated linear predictor of the modified and unmodified fast-GFPCA approaches; see Figure~\ref{fig:NHANES_corr_linear_predictor} and associated discussion. Moreover, $\hat{\beta}_0(s)$ and $\hat{\boldsymbol{\phi}}(s)$ are similar across the fast-GFPCA fits (overlapping vs non-overlapping windows, and bin widths), suggesting that the fast-GFPCA algorithm is fairly robust to the choice of input parameters in the NHANES data. 

However, there are substantial differences between vbFPCA and fast-GFPCA both in terms of the estimated population mean functions and first three eigenfunctions. The largest differences in $\hat{\beta}_0(s)$ occur around 6AM-8AM and 9PM-12AM, where vbFPCA respectively over- and under-estimates $\beta_0(s)$ relative to fast-GFPCA. The vbFPCA $\hat{\beta}_0(s)$ suggests that, for participants with $b_i(s) = 0$, the probability of being active between 6AM-8AM is around $80$-$90$\%  compared to around $50$\% from fast-GFPCA. This result is unexpected and does not match the results in the data, where we observe approximately $50$\% probability of being active during this time. One potential explanation could be that vbFPCA provides biased estimators of the mean and that the bias may be shifted into the latent random process variation. This hypothesis appears to be supported by our data driven simulation in Section~\ref{sec_supp:data_driven_sim} of the supplemental material. It is unclear at this time what is driving the observed bias in the vbFPCA approach as the same behavior is not seen in simulations presented in Section~\ref{sec:simulation}, but it may be related to the high proportion of observed $0$s (inactive periods) during the night time hours in our data application.

\paragraph{Linear Predictor.} Our goal in this section is to understand how differences in estimated model parameters $\hat{\beta}_0(s)$ and $\hat{\boldsymbol{\phi}}(s)$ across methods lead to differences in the estimated linear predictor $\hat{\eta}_i(s); i\in 1,\ldots, N$. To capture this, for each minute of the day we regressed $\hat{\eta_i}(s)$ from fast-GFPCA with non-overlapping windows and $w=6$ on $\hat{\eta_i}(s)$ from each of four other GFPCA models.  Figure~\ref{fig:NHANES_corr_linear_predictor} shows results of these time-specific linear regressions. These have the form $E[Y] = \mathcal{B}_0 + \mathcal{B}_1 X$, where $Y$ is $\hat{\eta_i}(s)$ from fast-GFPCA with no overlap windows and $w=6$, and $X$ is $\hat{\eta_i}(s)$ estimated by: (1) fast-GFPCA with non-overlapping windows and $w=30$ (orange lines); (2) modified fast-GFPCA with non-overlapping windows and $w=6$ (blue lines); (3) vbFPCA with Kt=8 (green lines); and (4) vbFPCA with Kt=30 (yellow lines). The first panel and second panels display regression coefficients $\hat{\mathcal{B}}_0$ and $\hat{\mathcal{B}}_1$, respectively, and the third panel displays the percent variance explained ($R^2$). The black dashed line corresponds to perfect agreement between $Y$ and $X$.

We observe almost perfect agreement between the reference model (fast-GFPCA with no overlap and $w = 6$) and the corresponding modified fast-GFPCA model (blue lines, intercept $\approx 0$, slope $\approx 1$, $R^2 \approx 1$, left to right panels, respectively). Similarly, we observe near perfect linear association between the fast-GFPCA model with larger bin width $w=30$ (orange lines) across the day ($R^2 \geq 0.95$, right panel), with a nearly 1-to-1 relationship (intercept $\approx 0$, slope $\approx 1$) during the active hours of the day ($\approx$ 10AM to 8PM). During the active hours of the day, the vbFPCA results suggest a similar 1-to-1 trend in the mean, though the lower $R^2$ suggests less agreement between the fast-GFPCA and vbFPCA approaches than across fast-GFPCA estimates with different input parameters.


\paragraph{Predictive Accuracy.} The observed differences in the estimated linear predictor do not appear to translate into different predictive accuracy in terms of either AUC of log loss; see the two rightmost columns of  Table~\ref{table:NHANES_results_ct_AUC_LL}. A possible explanation of the similar predictive accuracy but different estimation performance may be that the disagreement in predictions between models occurs primarily during the nighttime hours (Figure~\ref{fig:NHANES_corr_linear_predictor}), when the probability of an individual being active is generally very low. 

\paragraph{Computation Time.} Computation times for each model fit are presented in the middle of Table~\ref{table:NHANES_results_ct_AUC_LL} for fast-GFPCA (top rows) and the vbFPCA approach (bottom rows). For fast-GFPCA, computation times are broken down by step of the algorithm. The fast-GFPCA algorithm (top rows, ``Modified" = ``No") requires about $3$-$4$ hours while vbFPCA requires $7$ minutes for Kt=8 and $20$ minutes for Kt=30.  These computation times are driven by Step 4, which is unavoidable if the model is fit on the entire NHANES data set. Indeed, the fast-GFPCA approach here is the simplest GLMM that can be fit while maintaining the functional structure of the data. When Step 4 is modified (top rows, ``modified" = ``Yes") computation times decrease substantially and becomes comparable to the vbFPCA approach. From a practical perspective we have not found any substantial differences between the results obtained from the fast-GFPCA and modified fast-GFPCA approaches.


\section{Simulation Study}
\label{sec:simulation}

Our simulations are designed to (1) quantify the computational efficiency and scalability of fast-GFPCA as sample and grid size increase, (2) evaluate the accuracy of our method in comparison with existing approaches for generalized and binary FPCA, and (3) understand the behavior of our method under different data binning strategies from Step 1 of our estimation algorithm in Section \ref{subsec:methods_estimation}. For larger simulated datasets we focus on the binary functional data setting because the only existing competing approach that is computationally feasible for large functional datasets is tailored specifically to binary data. For smaller datasets we also evaluate our method in comparison to an existing approach for GFPCA  of Poisson functional data.

\subsection{Simulation Design}

We simulate binary functional data for $N = 100, 500, 1000$ subjects observed on a length $J = 100, 500, 2000$ grid in $S = [0, 1]$ that is equally spaced and shared across subjects. Poisson functional data are simulated from $N = 50, 100$ and $J = 100, 200$ due to the computational limitations of competing methods. Curves for $i\in 1,\ldots,N$ subjects are drawn from model~\eqref{eq:full_conditional}. We construct $K = 4$ principal components, with true eigenvalues $\lambda_k = 0.5^{k-1}; k = 1,2,3,4$. 
The true eigenfunctions are drawn from one of two scenarios intended to mimic real data settings. In the first setting, latent curves $\eta_i(s)$ and eigenfunctions are periodic, with $\boldsymbol{\phi}(s) = \{\sqrt{2}\sin(2\pi s), \sqrt{2}\cos(2\pi s), \sqrt{2}\sin(4\pi s), \sqrt{2}\cos(4\pi s)\}$. The second setting does not exhibit periodicity, with true eigenfunctions given by $\boldsymbol{\phi}(s) = \{1, \sqrt{3}(2s-1), \sqrt{5}(6s^2-6s + 1), \sqrt{7}(20s^3-30s^2+12s-1)\}$. 
For most simulation settings we assume $\beta_0(s) = 0$, however, for a subset of scenarios we construct nonzero $\beta_0(s)$ using a B-spline basis, specifically for binary functional data with: (1) $N, J = (1000, 2000)$; and (2) $N, J = (500, 100)$.

For binary and Poisson data, $g(\cdot)$ is taken to be the the logit and log links, respectively. For each subject and time point, exponential family observations $Z_i(s)$ are sampled independently from either a Bernoulli distribution with probability $logit^{-1}\left[\eta_i(s)\right]$, or from a Poisson distribution with rate $e^{\left[\eta_i(s)\right]}$.

\subsection{Comparison to Existing Approaches}

\subsubsection{Binary Functional Data}

We assess the performance of fast-GFPCA across different bin widths $w_l$ and compare non-overlapping and overlapping windows. Specifically, for each simulation scenario we evaluate three different bin widths, $w_l \in (6, 10, 50)$. We do not estimate fast-GFPCA with $w_l = 50$ when $J < 500$ as the large bin size does not make sense in this context.  
 
We compare fast-GFPCA with two different binary FPCA approaches, both of which are implemented in the {\ttfamily registr} package \cite{registr_package, registr2}. The first method to which we compare is the two-step conditional GFPCA model introduced by \cite{10.1016/j.csda.2016.07.010}, which is implemented using the {\ttfamily registr::gfpca\_twoStep()} function and referred to as \textit{tsGFPCA} in text and figures below. While {\ttfamily gfpca\_twoStep()} is a general purpose function that can accommodate multiple exponential family distributions, it is computationally intensive and thus impractical for most simulation settings. To reduce this computational burden we only implement {\ttfamily tsGFPCA} when $N \in (100, 500)$ and $J = 100$. We compare fast-GFPCA in all binary data simulation settings to the vbFPCA algorithm from \cite{10.1111/biom.12963}, which is implemented via the {\ttfamily registr::bfpca()} function, and is denoted {\ttfamily vbFPCA}. This method is highly computationally efficient, but designed for binary functional data with a logit link, and cannot be generalized to other link functions or exponential family distributions. Because the {\ttfamily vbFPCA} approach models population mean $\beta_0(s)$ and eigenfunctions $\boldsymbol{\phi(s)}$ using a B-spline expansion without a smoothness penalty, the number of basis functions must be manually tuned to obtain optimal smoothness. To address this, for each simulated dataset we implement {\ttfamily vbFPCA} with $Kt = 8$ basis functions (the package default) and $Kt = 30$.

For both competing methods we implement a periodic B-spline basis using the {\ttfamily registr} option {\ttfamily periodic = TRUE}. By default in the {\ttfamily registr} package, eigenfunctions $\boldsymbol{\phi(s)}$ are returned unscaled and on a grid of size 100. To enable comparison with results from fast-GFPCA, we linearly interpolate eigenfunctions estimated using \textit{tsGFPCA} and {\ttfamily vbFPCA} to a grid of size $J$ and scale by the square root of the grid length.

\subsubsection{Poisson Functional Data}

In the Poisson setting we compare fast-GFPCA with {\ttfamily registr::gfpca\_twoStep()}. Since the comparative method is highly computationally intensive, we only consider small data settings of sample sizes $N \in (50, 100)$, grid lengths $J \in (100, 200)$. We simulate 100 datasets for each of the six simulation scenarios arising from this combination of grid length and sample size. For fast-GFPCA we compare non-overlapping and overlapping windows, and consider bin widths $w_l \in (6, 10)$.

\subsection{Evaluation Criteria}

We compare the performance of the three methods ({\ttfamily tsGFPCA}, {\ttfamily vbFPCA}, and {\ttfamily fastGFPCA}) with respect to accuracy in recovering subject-specific latent means in the linear predictor space $\eta_i(s)$, accuracy in recovering  population-level mean $\beta_0(s)$ and eigenfunctions $\boldsymbol{\phi(s)}$, and computational efficiency. Accuracy of subject-specific log-odds across models is quantified using mean integrated squared error (MISE) across subjects, given by $\frac{1}{N}\sum_{i - 1}^N \int_0^1\left(\hat{\eta}_i(s)-\eta_i(s) \right)^2ds$. Accuracy of eigenfunction estimation is compared using MISE defined by $\frac{1}{k}\sum_{k=1}^4\int_0^1\left(\hat{\phi}_k(s)-\phi_k(s) \right)^2ds$, and population mean accuracy is measured using ISE. Computation times are reported in minutes.

\subsection{Simulation Results: Accuracy}

Tables \ref{table:etaMISE}, \ref{table:phiMISE}, and \ref{table:muISE} summarize accuracy of key quantities across methods and simulation scenarios, where outcome data were generated from both binomial and Poisson distributions with periodic true eigenfunctions. Table \ref{table:etaMISE} provides the MISE for $\hat{\eta}_i(s)$, the estimated subject-specific latent means in the linear predictor space, which are log-odds for binomial data and log-rates for Poisson data. Similarly, Tables \ref{table:phiMISE} and \ref{table:muISE} summarize the MISE of eigenfunctions $\boldsymbol{\phi}(s)$ and ISE of population-level mean $\beta_0(s)$, respectively.

\subsubsection{Binomial data}

Table \ref{table:etaMISE} shows that all methods estimate the latent log-odds accurately. Our {\ttfamily fastGFPCA} approach performs equally well to the best competing method, {\ttfamily vbFPCA} with $Kt = 8$ spline bases in all but one scenario. Simulated data are periodic, and as a result {\ttfamily fastGFPCA} with overlapping bins outperforms {\ttfamily fastGFPCA} with non-overlapping bins. {\ttfamily fastGFPCA} performs best for the largest bin width, $w_l =50$, except when the grid size $J$ is smallest ($J = 100$). However, {\ttfamily fastGFPCA} performs well across all chosen bin widths. 

Table \ref{table:phiMISE} indicates that our {\ttfamily fastGFPCA} approach recovers true population eigenfunctions comparably to or better than the competing {\ttfamily vbFPCA} method in every scenario. For smaller grid sizes $J$ the {\ttfamily vbFPCA} approach with with $Kt = 8$ performs slightly better, but {\ttfamily fastGFPCA} outperforms {\ttfamily vbFPCA} when $J$ increases. Table \ref{table:muISE} shows that {\ttfamily fastGFPCA} outperforms {\ttfamily vbFPCA} at recovering the population mean $\beta_0(s)$ in all simulation scenarios. This is likely due to the fact that {\ttfamily fastGFPCA} penalizes spline coefficients to obtain a flat line estimate around the true value of  $\beta_0(s) = 0$ while {\ttfamily vbFPCA}, which does not penalize smoothness, cannot estimate $\beta_0(s)$ as well when the true function is linear. Similar results for non-periodic data are observed in Supplemental Table \ref{table:supp_case2}.

Figure \ref{fig:simulation_accuracy} highlights these results at a more granular level for one simulation scenario with $N = 500$ subjects and $J = 500$ time points and a nonzero, nonlinear population mean function $\beta_0(t)$. Specifically, Figure \ref{fig:simulation_accuracy} shows
the estimated population mean function, $\hat{\beta}_0(s)$, and the first four estimated eigenfunctions, $\hat{\phi}_k(s); k \in 1,\ldots,4$, from 100 simulated datasets. Estimates are presented as red lines for models using the competing {\ttfamily vbFPCA} approach or blue lines for models using our proposed {\ttfamily fastGFPCA} method, with dotted black lines representing the true value. All methods provide reasonable results in this simulation setting. Our {\ttfamily fastGFPCA} approach with overlapping bins and bin width $w_l =50$ provides the best results. The {\ttfamily vbFPCA} method with $Kt = 8$ spline basis functions also performs well, though overestimates $\beta_0(t)$ at the beginning of the functional domain $s$.  The {\ttfamily fastGFPCA} approach with non-overlapping bins and $w_l =50$ overestimates $\phi_4(s)$ at the endpoints of the functional domain, which suggests that a smaller band width may be more appropriate when data is not periodic and only non-overlapping bins can be used for {\ttfamily fastGFPCA}.

\subsubsection{Poisson data}

For Poisson distributed functional data the proposed {\ttfamily fastGFPCA} method was compared with the two-step approach {\ttfamily tsGFPCA} because there is no variational EM method for Poisson FPCA. Table \ref{table:phiMISE} indicates that {\ttfamily fastGFPCA} estimates the latent log-rate far better than the competing {\ttfamily 
 tsGFPCA} method in every simulation scenario.  Similarly, Tables \ref{table:phiMISE} and \ref{table:muISE} indicate that {\ttfamily fastGFPCA} recovers the true population mean and eigenfunctions much more accurately than {\ttfamily tsGFPCA} in every scenario. Of the {\ttfamily fastGFPCA} approaches, {\ttfamily fastGFPCA} with overlapping bins and and $w_l =10$ tends to perform best, but the different {\ttfamily fastGFPCA} models perform similarly regardless of window overlap or choice of bin width $w_l$.

Figure \ref{fig:simulation_accuracy_poisson} in the supplemental material provides some intuition as to why {\ttfamily tsGFPCA} performs poorly in the Poisson setting. This figure shows the estimated population mean function and eigenfunctions from 25 simulated datasets with $N = 100$ subjects and $J = 100$ time points. Model estimates are presented as red lines ({\ttfamily tsGFPCA}) or blue lines ({\ttfamily fastGFPCA}). The proposed {\ttfamily fastGFPCA} method provides reasonable results for bin widths $w_l =6$ and $w_l =10$, but {\ttfamily tsGFPCA} clearly provides incorrect estimates.

\subsection{Simulation Results: Computational Efficiency}

Table \ref{table:comptime} shows median computation time in minutes across methods and simulation scenarios for both binomial and Poisson functional data. Across all scenarios the non-overlapping {\ttfamily fastGFPCA} approach is more computationally efficient than the overlapping {\ttfamily fastGFPCA} approach. Bin widths $w_l$ have a negligible effect on computation time for non-overlapping bins, but when bins are overlapping computation time increases with increasing bin width. {\ttfamily fastGFPCA} scales well when grid size $J$ increases but more slowly when one increases the number of subjects, $N$. Notably, at smaller sample sizes ($N \in \{100, 500\})$, {\ttfamily fastGFPCA} is comparably efficient or faster than {\ttfamily vbFPCA}, a method custom-built for speed. For $N \ge 1000$, {\ttfamily fastGFPCA} can be sped-up using techniques discussed in Supplemental Section \ref{sec_supp:speed_up}. The \textit{tsGFPCA} has a median time of 92 minutes for Poisson data with just 100 subjects and 200 time points. This indicates that  \textit{tsGFPCA} 
is prohibitively slow for our data application of $4286$ subjects with 1440 time points each.


\section{Discussion}
\label{sec:discussion}

The fast-GFPCA method proposed in this manuscript represents a simple, understandable, and computationally feasible solution to the complex task of estimating functional principal components analysis for non-Gaussian data. In addition, we have provided a mathematical justification for the principles that underlie fast-GFPCA and shown that the method compares favorably to the few existing approaches in terms of both estimation accuracy and computational efficiency. Moreover, existing methods, such as the vbFPCA approach for binary data presented here may provide biased estimates of model parameters (population mean function and covariance operator), suggesting that the fast-GFPCA approach is a reasonable method for comparison even when an alternative approach is more computationally efficient in a given application.

Though the work here shows fast-GFPCA to be fast, accurate, and appropriate for analyzing the motivating NHANES data, methodologic work remains. Specifically, it is unclear at this time how to choose the optimal bin, both with regard to bin width and the decision to use overlapping versus non-overlapping windows for estimation. While a cross-validated prediction error criteria may be a viable option, subject-level cross-validation requires prediction of random effects in non-Gaussian models using participants' data not included in model fitting, a non-trivial problem in non-Bayesian contexts. Moreover, when using non-overlapping windows, automated smoothing parameter selection of FPCA on the latent process in Step 3 of the fast-GFPCA algorithm is unreliable. Here we propose an ad-hoc solution based on visual inspection of the eigenfunctions and/or estimated covariance function. While this is feasible due to the speed of the FACE method implemented in Step 3, we would prefer a fully automated approach. Deriving an appropriate variation of the GCV criteria used by FACE for smoothing parameter selection which accounts for autocorrelated data may improve the method proposed here. Nevertheless, the results of this work represent an encouraging step forward for estimation of GFPCA in very high dimensional data, specifically large $N$, a key bottleneck for the application of functional data analysis methods in practice. 

An appealing feature of fast-GFPCA is that it can be extended to: (1) covariate dependent GFPCA; (2) multilevel, structured, and longitudinal GFPCA. 

\paragraph{Covariate dependent GFPCA.} The fast-GFPCA method easily incorporates covariates into the model. Consider the case of one additional scalar predictor (e.g., age), denoted $x_i$. Step 2 of fast-GFPCA is simply modified  to fit local models of the form $g(E[Z_i(s_j)| s_j \in S_l]) = \beta_{0}(s_{m_l}) + \beta_1(s_{m_l})x_i + b_i(s_{m_l}) = \eta_i(s_{m_l})$. Then, in Step 4, the final GLMM includes the additive term associated with the proposed varying coefficient model. The effort required for this extension is minimal. 

\paragraph{Nested, longitudinal or crossed design GFPCA.} Consider the case when multiple functions are observed per study participant. For example, in the NHANES data each study participant has  multiple days of accelerometry data. For notation simplicity assume that there are $K$ functions for every study participant. The extension to multilevel GFPCA follows naturally from the fast-GFPCA algorithm. Specifically, in Step 2 we fit the multilevel model $g(E[Z_{ik}(s_j)]) = \beta_0(s_{m_l}) + b_i(s_{m_l}) + v_{ik}(s_{m_l})$. Step 3 can then proceed with MFPCA FACE \cite{10.1080/10618600.2022.2115500} to estimate the principal directions of variation at each level. If the functional data has  longitudinal \cite{greven2010longitudinal} or crossed \cite{shou2015} designs the local GLMM can be changed accordingly.


\singlespacing
\bibliography{main}

\clearpage

\begin{figure}[!ht]
\centering
\includegraphics[width=0.8\textwidth]{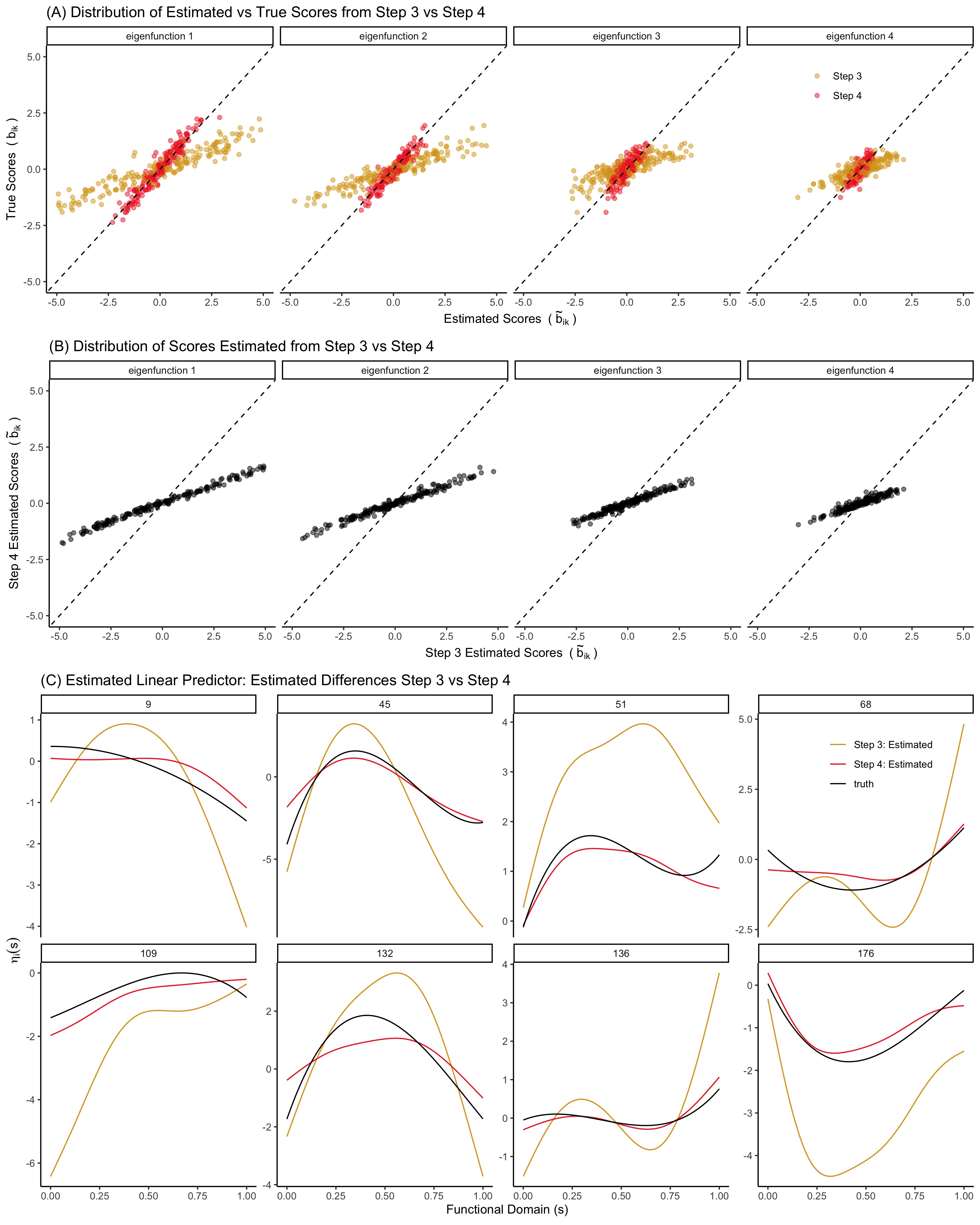}
\caption{Illustration of the need for Step 4 in the fast-GFPCA method using one simulated dataset from the simulation study ($N=200$, $J=200$, $w_l=10$, $\boldsymbol{\phi}(s) = \{\sqrt{2}\sin(2\pi s), \sqrt{2}\cos(2\pi s), \sqrt{2}\sin(4\pi s), \sqrt{2}\cos(4\pi s)\}$).  (A) Plot of the estimated scores from step 3 (golden points) and step 4 (red points) on the x-axis versus the true scores on the y-axis separately for each of the first four eigenfunctions. The black line represents the identity line. (B) Plot of the scores estimated from step 3 (x-axis) versus those from step 4 (y-axis). (C) Estimated curves on the linear predictor scale}
\label{fig:need_for_step_4}
\end{figure}

\begin{figure}[!ht]
    \centering
    \includegraphics[width=\textwidth]{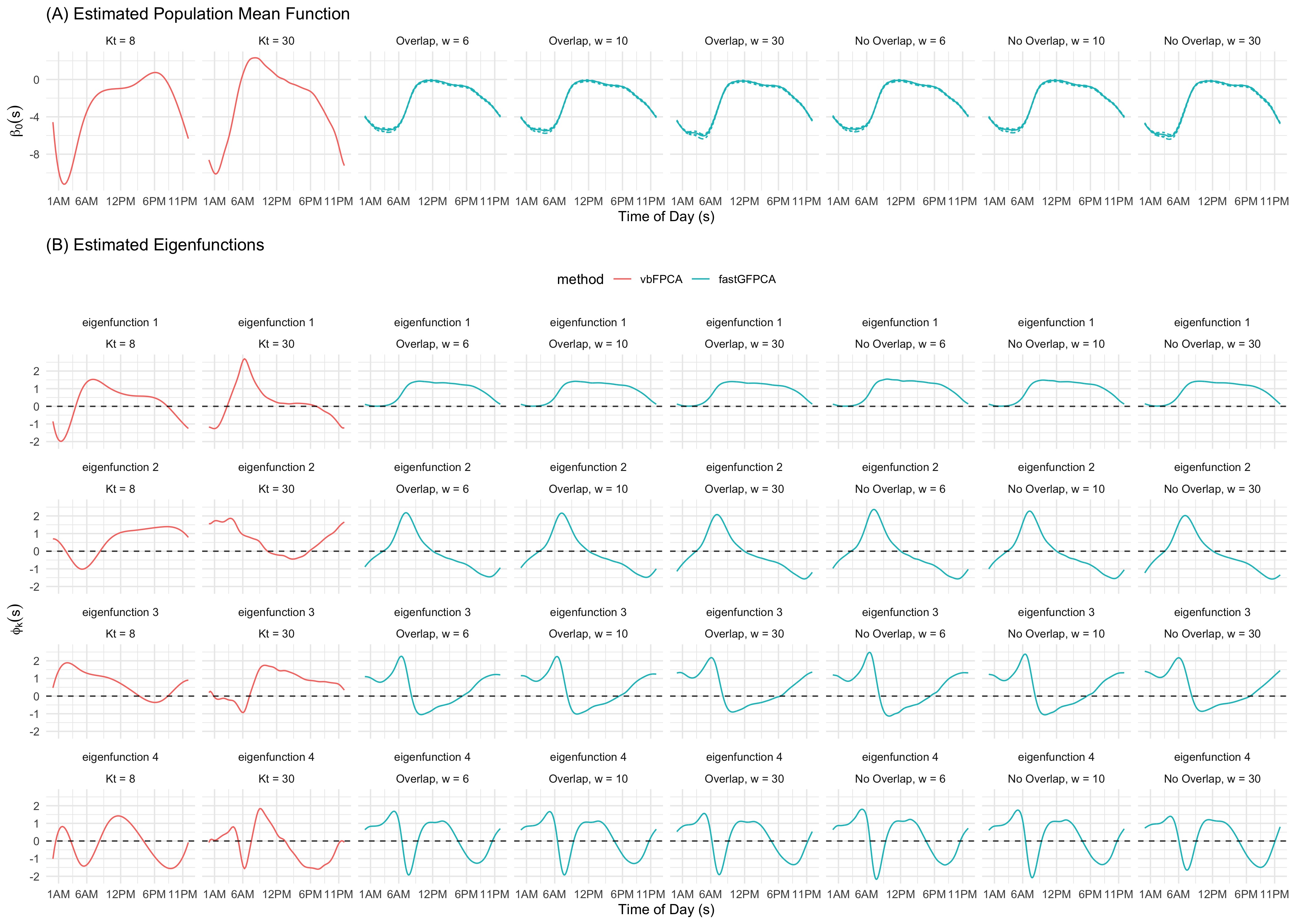}
    \caption{
    Estimated population mean function (first row) and the first four estimated eigenfunctions (rows $2-5$) in NHANES. Model estimates are presented as red (vbFPCA) or blue lines (fast-GFPCA). The two leftmost columns correspond to vbFPCA (Kt=8 in column 1 and Kt=30 in column 2). The six rightmost columns correspond to fast-GFPCA with different input parameters (overlapping versus non-overlapping windows) and window sizes ($w= 6, 10, 30$). Estimates of the population mean function based on the modified fast-GFPCA are displayed as dashed lines.
    \label{fig:NHANES_results_estimated_parameters}
    }
\end{figure}

\begin{figure}[!ht]
    \centering
    \includegraphics[width=\textwidth]{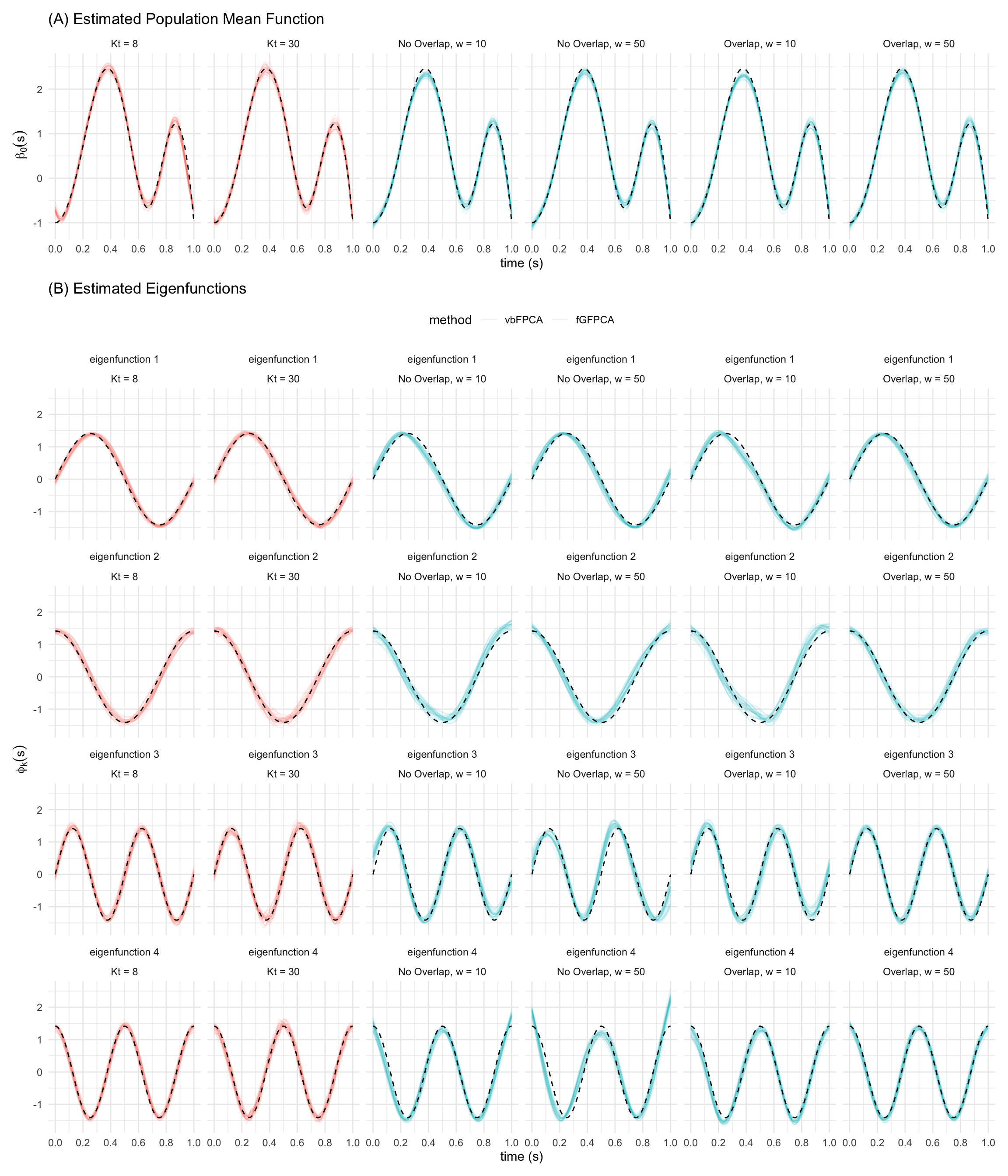}
    \caption{The estimated population mean function $\hat{\beta}_0(s)$ and eigenfunctions $\hat{\boldsymbol\phi}(s)$ from 100 simulated binary functional datasets with $N = 500$ subjects and $J = 500$ time points. Model estimates are red lines ({\ttfamily vbFPCA}) or blue lines ({\ttfamily fastGFPCA}). {\ttfamily vbFPCA} estimates from models using either $Kt = 8$ (left column) or $Kt  = 30$ (second column) basis functions are compared with six {\ttfamily fastGFPCA} models. From left to right, the {\ttfamily fastGFPCA} models (blue columns) are estimated with non-overlapping windows with bin sizes $w_l =10$ and $w_l =50$, and overlapping windows with bin sizes $w_l =10$ and $w_l =50$.
    \label{fig:simulation_accuracy}
    }
\end{figure}


\begin{table}[!ht]
    \centering
    \scalebox{0.85}{
    \begin{tabular}{l rrrrrr r r}\toprule
    \multicolumn{9}{c}{{\ttfamily fastGFPCA}} \\ \cmidrule(lr){1-9}
    Parameters & Modified & \multicolumn{5}{c}{Computation Time (mins)} & \multicolumn{1}{c}{AUC} & \multicolumn{1}{c}{Log-loss} \\ \cmidrule(lr){1-1} \cmidrule(lr){2-2}\cmidrule(lr){3-7}\cmidrule(lr){8-8}\cmidrule(lr){9-9}
     && Step 1 & Step 2 & Step 3 & Step 4 & Total & & \\\cmidrule(lr){3-3}\cmidrule(lr){4-4}\cmidrule(lr){5-5}\cmidrule(lr){6-6}\cmidrule(lr){7-7}
    \multirow{2}{*}{Overlap, $w=6$} & No  & & 16.42 & 0.04 & 174.20 & 190.63 & 0.909 & 0.328 \\ 
                                    & Yes & & 16.42 & 0.04 & 10.90 & 27.33 & 0.909 & 0.328 \\ 
    \multirow{2}{*}{Overlap, $w=10$} & No  & & 23.28 & 0.04 & 184.82 & 208.12 & 0.909& 0.327 \\ 
                                    & Yes &  & 23.28 & 0.04 & 10.74 & 34.03 & 0.909 & 0.327\\ 
    \multirow{2}{*}{Overlap, $w=30$} & No  & & 55.90 & 0.04 & 184.66 & 240.57 & 0.909 & 0.327\\ 
                                    & Yes & & 55.90 & 0.04 & 10.48  & 66.38 & 0.909 & 0.327 \\ 
    \multirow{2}{*}{No Overlap, $w=6$} & No  & & 2.48 & 0.01 & 188.90 & 191.43 & 0.909 & 0.328 \\ 
                                       & Yes & & 2.48 & 0.01 & 11.34 & 13.13 & 0.909 & 0.328 \\ 
    \multirow{2}{*}{No Overlap, $w=10$} & No  & & 2.19 & 0.01 & 185.04 & 187.28 & 0.909 & 0.327\\ 
                                       & Yes & & 2.19 & 0.01 & 10.89 & 13.13 & 0.909 & 0.327\\ 
    \multirow{2}{*}{No Overlap, $w=30$} & No  && 1.97 & 0.01 & 197.35 & 199.36 & 0.909  & 0.327\\ 
                                       & Yes & & 1.97 & 0.01 & 11.60 & 13.61 & 0.909 & 0.327 \\ \midrule
    \multicolumn{9}{c}{{\ttfamily vbFPCA} ({\ttfamily registr::bfpca()}) } \\ \cmidrule(lr){1-9}
    Parameters & & \multicolumn{5}{c}{Computation Time (mins)} & \multicolumn{1}{c}{AUC} & \multicolumn{1}{c}{Log-loss} \\ \cmidrule(lr){1-1} \cmidrule(lr){3-7}\cmidrule(lr){8-8}\cmidrule(lr){9-9}
    Kt=8 & &\multicolumn{5}{r}{7.11} & 0.909 & 0.328 \\ 
    Kt=30 & & \multicolumn{5}{r}{20.86} & 0.910 & 0.326 \\ \bottomrule
    \end{tabular}
    }
    \caption{
    Computation times and in-sample predictive performance summaries for {\ttfamily fastGFPCA} (top rows) estimated under different parameter settings (column 1, overlapping versus non-overlapping windows, window size of $w=6, 10, 30$) using both the primary and modified algorithm (column 2, Modified = No or Yes, respectively), and variational Bayes ({\ttfamily vbFPCA}, bottom rows) estimated with either $Kt=8$ or $Kt=30$. For {\ttfamily fastGFPCA}, computation times are presented using both the total time and separately by each step of the procedure (step 1 is left blank as computationally this step is effectively instantaneous). AUC and log-loss, presented in the rightmost two columns are calculated over all minutes of the day.
    \label{table:NHANES_results_ct_AUC_LL}
    }
\end{table}

\begin{table}[!ht]
\scalebox{0.75}{
\begin{tabular}{@{}llllllllllll@{}}\toprule
\multicolumn{12}{c}{Mean Integrated Squared Error of $\eta_i(s)$}                                                                                                                                                                       \\\cmidrule(lr){1-12}
                        &                    &                    & \multicolumn{3}{l}{Fast-GFPCA}                               &           &           &                  & vbFPCA        &               & tsGFPCA \\ \cmidrule(lr){4-9} \cmidrule(lr){10-11}\cmidrule(lr){12-12}
\multirow{2}{*}{Family} & \multirow{2}{*}{N} & \multirow{2}{*}{J} & \multicolumn{3}{l}{Overlapping Bins}       & \multicolumn{3}{l}{Non-Overlapping Bins} & K = 8         & K = 30        & K = 30  \\
                        &                    &                    & $w_l =6$      & $w_l =10$        & $w_l =50$        & $w_l =6$     & $w_l =10$    & $w_l =50$           &               &               &         \\\cmidrule(lr){1-12}
Binomial                & 100                & 100                & 2.19       & \textbf{2.14} & -             & 2.39      & 2.44      & -                & 2.15          & 2.87          & 2.18    \\
                        &                    & 500                & 0.55       & 0.52          & \textbf{0.49} & 0.53      & 0.53      & 0.58             & \textbf{0.49} & 0.63          & -       \\
                        &                    & 2000               & 0.17       & 0.15          & \textbf{0.13} & 0.15      & 0.15      & \textbf{0.13}    & \textbf{0.13} & 0.16          & -       \\
                        & 500                & 100                & 2.05       & \textbf{2.02} & -             & 2.33      & 2.43      & -                & \textbf{2.02} & 2.15          & 2.03    \\
                        &                    & 500                & 0.5        & 0.49          & \textbf{0.47} & 0.5       & 0.5       & 0.58             & \textbf{0.47} & 0.5           & -       \\
                        &                    & 2000               & 0.15       & 0.14          & \textbf{0.12} & 0.14      & 0.14      & 0.13             & \textbf{0.12} & 0.13          & -       \\
                        & 1000               & 100                & 2.05       & 2.04          & -             & 2.33      & 2.41      & -                & \textbf{2.01} & 2.08          & -       \\
                        &                    & 500                & 0.5        & 0.48          & \textbf{0.47} & 0.51      & 0.5       & 0.58             & \textbf{0.47} & 0.48          & -       \\
                        &                    & 2000               & 0.14       & 0.13          & \textbf{0.12} & 0.14      & 0.13      & 0.13             & \textbf{0.12} & \textbf{0.12} & -       \\\cmidrule(lr){1-12}
Poisson                 & 50                 & 100                & 0.52       & \textbf{0.49} & -             & 0.56      & 0.65      & -                & -             & -             & 13.0   \\
                        &                    & 200                & 0.32       & \textbf{0.28} & -             & 0.32      & 0.29      & -                & -             & -             & 15.0   \\
                        & 100                & 100                & 0.45       & \textbf{0.43} & -             & 0.5       & 0.59      & -                & -             & -             & 15.9    \\
                        &                    & 200                & 0.25       & \textbf{0.24} & -             & 0.27      & 0.25      & -                & -             & -             & 20.5  \\
\bottomrule\end{tabular}
}
    \caption{Mean integrated squared error (MISE) for $\hat{\eta}_i(s)$, the estimated subject-specific latent means in the linear predictor space across methods and simulation scenarios. In each row the method(s) with the lowest MISE for that simulation scenario is in \textbf{bold}. All data summarized in this table were simulated with population mean $\beta_0(s) = 0$. An ``-'' indicates that that model was not evaluated for a given simulation scenario. All values are multiplied by a factor of $10$.
    \label{table:etaMISE}
    }
\end{table}

\begin{table}[!ht]
\scalebox{0.75}{
\begin{tabular}{llllllllllll}\toprule
\multicolumn{12}{c}{Mean Integrated Squared Error of Eigenfunctions $\boldsymbol{\phi}(s)$}                                                                                                                                                                   \\\cmidrule(lr){1-12}
                        &                    &                    & \multicolumn{3}{l}{Fast-GFPCA}                          &           &           &                  & vbFPCA        &        & tsGFPCA \\\cmidrule(lr){4-9} \cmidrule(lr){10-11}\cmidrule(lr){12-12}
\multirow{2}{*}{Family} & \multirow{2}{*}{N} & \multirow{2}{*}{J} & \multicolumn{3}{l}{Overlapping Bins}       & \multicolumn{3}{l}{Non-Overlapping Bins} & K = 8         & K = 30 & K = 30  \\
                        &                    &                    & $w_l =6$      & $w_l =10$        & $w_l =50$        & $w_l =6$     & $w_l =10$    & $w_l =50$           &               &        &         \\\cmidrule(lr){1-12}
Binomial                & 100                & 100                & 0.91       & 0.66          & -             & 1.13      & 1.56      & -                & \textbf{0.47} & 1.13   & 0.48    \\
                        &                    & 500                & 0.66       & 0.55          & 0.46          & 0.57      & 0.59      & 0.58             & \textbf{0.28} & 0.39   & -       \\
                        &                    & 2000               & 0.64       & 0.54          & \textbf{0.42} & 0.55      & 0.52      & \textbf{0.42}    & 0.72          & 0.88   & -       \\
                        & 500                & 100                & 0.19       & 0.18          & -             & 0.58      & 0.8       & -                & \textbf{0.12} & 0.26   & 0.13    \\
                        &                    & 500                & 0.13       & 0.11          & \textbf{0.08} & 0.13      & 0.11      & 0.32             & \textbf{0.08} & 0.11   & -       \\
                        &                    & 2000               & 0.14       & 0.12          & \textbf{0.1}  & 0.13      & 0.13      & \textbf{0.1}     & 0.56          & 0.62   & -       \\
                        & 1000               & 100                & 0.11       & 0.08          & -             & 0.5       & 0.7       & -                & \textbf{0.05} & 0.11   & -       \\
                        &                    & 500                & 0.06       & 0.04          & \textbf{0.03} & 0.07      & 0.07      & 0.26             & 0.05          & 0.07   & -       \\
                        &                    & 2000               & 0.05       & 0.04          & \textbf{0.03} & 0.05      & 0.04      & 0.04             & 0.52          & 0.66   & -       \\\cmidrule(lr){1-12}
Poisson                 & 50                 & 100                & 1.03       & \textbf{0.94} & -             & 1.16      & 1.24      & -                & -             & -      & 10.97   \\
                        &                    & 200                & 0.96       & \textbf{0.88} & -             & 0.91      & 0.98      & -                & -             & -      & 8.8     \\
                        & 100                & 100                & 0.4        & \textbf{0.34} & -             & 0.52      & 0.66      & -                & -             & -      & 10.79   \\
                        &                    & 200                & 0.37       & \textbf{0.34} & -             & 0.4       & 0.41      & -                & -             & -      & 8.7\\   
\bottomrule\end{tabular}
}
    \caption{Mean integrated squared error (MISE) for estimated population-level latent eigenfunctions, $\phi_k(s); k = 1,\ldots, 4$. In each row the method(s) with the lowest MISE for that simulation scenario is in \textbf{bold}. All data summarized in this table were simulated with population mean $\beta_0(s) = 0$. An ``-'' indicates that that model was not evaluated for a given simulation scenario. All values are multiplied by a factor of $10$.
    \label{table:phiMISE}
    }
\end{table}

\begin{table}[!ht]
\scalebox{0.75}{
\begin{tabular}{llllllllllll}\toprule
\multicolumn{12}{c}{Integrated Squared Error of Population Mean $\beta_0(s)$}                                                                                                                                                                    \\\cmidrule(lr){1-12}
                        &                    &                    & \multicolumn{3}{l}{Fast-GFPCA}                          &               &               &               & vbFPCA &        & tsGFPCA \\ \cmidrule(lr){4-9} \cmidrule(lr){10-11}\cmidrule(lr){12-12}
\multirow{2}{*}{Family} & \multirow{2}{*}{N} & \multirow{2}{*}{J} & \multicolumn{3}{l}{Overlapping Bins}          & \multicolumn{3}{l}{Non-Overlapping Bins}      & K = 8  & K = 30 & K = 30  \\
                        &                    &                    & $w_l =6$         & $w_l =10$        & $w_l =50$        & $w_l =6$         & $w_l =10$        & $w_l =50$        &        &        &         \\\cmidrule(lr){1-12}
Binomial                & 100                & 100                & \textbf{1.34} & 1.35          & -             & 2.42          & 1.5           & -             & 15.69  & 27.37  & 1.85    \\
                        &                    & 500                & 0.53          & 0.54          & 0.48          & \textbf{0.42} & \textbf{0.42} & 0.62          & 12.94  & 15.88  & -       \\
                        &                    & 2000               & 0.1           & 0.09          & 0.11          & \textbf{0.06} & 0.08          & 0.07          & 12.13  & 12.11  & -       \\
                        & 500                & 100                & 0.19          & \textbf{0.17} & -             & 0.3           & 0.41          & -             & 3.53   & 5.74   & 0.46    \\
                        &                    & 500                & 0.06          & 0.06          & 0.06          & \textbf{0.05} & \textbf{0.05} & 0.06          & 3.03   & 3.45   & -       \\
                        &                    & 2000               & 0.02          & 0.02          & \textbf{0.01} & 0.02          & 0.02          & 0.02          & 3.12   & 3.13   & -       \\
                        & 1000               & 100                & 0.1           & \textbf{0.09} & -             & 0.14          & 0.12          & -             & 1.69   & 2.87   & -       \\
                        &                    & 500                & 0.04          & 0.04          & 0.04          & \textbf{0.03} & 0.05          & 0.04          & 1.57   & 1.77   & -       \\
                        &                    & 2000               & 0.02          & 0.02          & 0.02          & 0.02          & 0.02          & \textbf{0.01} & 2.06   & 1.87   & -       \\\cmidrule(lr){1-12}
Poisson                 & 50                 & 100                & 4.34          & \textbf{3.46} & -             & 5.12          & 5.54          & -             & -      & -      & 116.59  \\
                        &                    & 200                & 2.43          & \textbf{2.18} & -             & 3.68          & 3.45          & -             & -      & -      & 126.52  \\
                        & 100                & 100                & 2.31          & \textbf{2.18} & -             & 3.27          & 4.81          & -             & -      & -      & 174.8   \\
                        &                    & 200                & \textbf{1.64} & 1.67          & -             & 4.1           & 2.36          & -             & -      & -      & 148.47\\ 
\bottomrule\end{tabular}
}
    \caption{Integrated squared error (ISE) for estimated population-level latent mean, $\beta_0(s)$.  In each row the method(s) with the lowest ISE for that simulation scenario is in \textbf{bold}. All data summarized in this table were simulated with population mean $\beta_0(s) = 0$. An ``-'' indicates that that model was not evaluated for a given simulation scenario. All values are multiplied by a factor of $10^{3}$.
    \label{table:muISE}
    }
\end{table}

\begin{table}[!ht]
\scalebox{0.75}{
\begin{tabular}{@{}llllllllllll@{}}\toprule
\multicolumn{12}{c}{Median Computation Times}                                                                                                                                                    \\\cmidrule(lr){1-12}
                        &                    &                    & Fast-GFPCA    &           &          &               &               &              & vbFPCA       &        & tsGFPCA \\ \cmidrule(lr){4-9} \cmidrule(lr){10-11}\cmidrule(lr){12-12}
\multirow{2}{*}{Family} & \multirow{2}{*}{N} & \multirow{2}{*}{J} & \multicolumn{3}{l}{Overlapping Bins} & \multicolumn{3}{l}{Non-Overlapping Bins}     & K = 8        & K = 30 & K = 30  \\ 
                        &                    &                    & $w_l =6$         & $w_l =10$    & $w_l =50$   & $w_l =6$         & $w_l =10$        & $w_l =50$       &              &        &         \\\cmidrule(lr){1-12}
Binomial                & 100                & 100                & 0.1           & 0.1       & -        & 0.1           & \textbf{0}             & -            & \textbf{0}   & 0.2    & 16.9    \\
                        &                    & 500                & 0.5           & 0.5       & 1.6      & \textbf{0.1}  & \textbf{0.1}  & \textbf{0.1} & \textbf{0.1}          & 0.7    & -       \\
                        &                    & 2000               & 2             & 2.2       & 6.5      & 0.3           & \textbf{0.2}  & \textbf{0.2} & 0.9          & 4.7    & -       \\
                        & 500                & 100                & 2.3           & 2.3       & -        & 2.2           & 2.2           & -            & \textbf{0.1} & 0.7    & 71.3    \\
                        &                    & 500                & 2.7           & 3.2       & 7        & 2             & 1.9           & 1.9          & \textbf{0.6} & 2.8    & -       \\
                        &                    & 2000               & 7.4           & 9.5       & 27       & \textbf{3.1}  & \textbf{3.1}  & \textbf{3.1} & 5.7          & 24.4   & -       \\
                        & 1000               & 100                & 20.4          & 20.1      & -        & 20.2          & 20.4          & -            & \textbf{0.4} & 1.3    & -       \\
                        &                    & 500                & 18.6          & 20        & 29.9     & 17.2          & 17.4          & 17.2         & \textbf{1.7} & 6.4    & -       \\
                        &                    & 2000               & 30.1          & 35.1      & 75.9     & 22.3          & 22.6          & 22.4         & \textbf{21}  & 71.4   & -       \\\cmidrule(lr){1-12}
Poisson                 & 50                 & 100                & 0.1           & 0.1       & -        & \textbf{0.02} & \textbf{0.02} & -            & -            & -      & 11.2    \\
                        &                    & 200                & 0.2           & 0.21      & -        & 0.04          & \textbf{0.03} & -            & -            & -      & 40.4    \\
                        & 100                & 100                & 0.15          & 0.16      & -        & \textbf{0.06} & \textbf{0.06} & -            & -            & -      & 34.6    \\
                        &                    & 200                & 0.25          & 0.27      & -        & 0.07          & \textbf{0.06} & -            & -            & -      & 92.0   \\
\bottomrule\end{tabular}
}
    \caption{Median computation time in minutes across methods and simulation scenarios. In each row the method(s) with the fastest computation time for that simulation scenario is in \textbf{bold}. All data summarized in this table were simulated with population mean $\beta_0(s) = 0$. An ``-'' indicates that that model was not evaluated for a given simulation scenario. 
    \label{table:comptime}
    }
\end{table}

\clearpage
\doublespacing

\section*{Supplemental Material}
\setcounter{page}{1}
\setcounter{section}{0}
\setcounter{table}{0}
\setcounter{figure}{0}

\renewcommand{\thesection}{S-\arabic{section}}
\renewcommand{\thetable}{S\arabic{table}}
\renewcommand{\thefigure}{S\arabic{figure}}

\section{Algorithmic Presentation of fast-GFPCA}
\label{sec_supp:algorithm}
\small
\begin{algorithm}[H]
\caption{Fast-GFPCA}\label{alg:fastGFPCA}
\stepone{}{
    \Input{$[\{Z_i(s_j) \}, 1 \leq i \leq N, 1 \leq j \leq J]$}
    \Output{$[\{Z_i(s_j), l\} 1 \leq i \leq N, j \in \mathcal{S}_l, 1 \leq l \leq L  \}]$}
}
\steptwo{}{
    \Input{$[\{Z_i(s_j), l\} 1 \leq i \leq N, j \in \mathcal{S}_l, 1 \leq l \leq L  \}]$}
    \Output{$[\{\hat{\eta}_i(s_{m_l}), l\}, 1 \leq i \leq N, 1 \leq l \leq L]$}
    \For{$l = 1,\ldots, L$}{
        Fit a GLMM of the form: 
            $$g(E[Z_i(s_j)| s_j \in S_l]) = \beta_{0}(s_{m_l}) + b_i(s_{m_l}) = \eta_i(s_{m_l})$$
        Obtain $\widehat{\eta}_i(s_{m_l}) = \hat{\beta}_0(s_{m_l}) + \widehat{b}_i(s_{m_l})$ 
    }
}
\stepthree{}{
    \Input{$[\{\widehat{\eta}_i(s_{m_l}), l\}, 1 \leq i \leq N, 1 \leq l \leq L]$}
    \Output{$[\{\hat{\phi}_k(s_{m_l}), l\}, 1 \leq l \leq L, 1 \leq k \leq K]$}
    Estimate $\text{Cov}\{\widehat{\eta}_i(u), \widehat{\eta}_i(v)\}$ for $u,v = m_1,\ldots, m_L$ using the FACE algorithm \\
    Obtain $\hat{\phi}_k(s)$ for $s = m_1,\ldots, m_L$
    
}
\stepfour{}{
    \Input{$[\{\hat{\phi}_k(s_l), l\}, 1 \leq l \leq L, 1 \leq k \leq K],  [\{Z_i(s_j) \}, 1 \leq i \leq N, 1 \leq j \leq J]$}
    \Output{$[\{\widehat{b}_i(s_j), \hat{\beta}_0(s_j)\}, 1 \leq i \leq N, 1 \leq j \leq J]$}
    \For{$k = 1, \ldots, K$}{
        Obtain $\widehat{\phi}_k(s)$ on the original grid $s_1,\ldots, s_J$ 
    }
    Estimate GFPCA using additive generalized linear mixed models of the form 
    \begin{align*}
        g(E[Z_i(s)]|\{\widehat{\phi}_k(s): 1 \leq s \leq J, 1 \leq k \leq K \}) &= \beta_0(s) + b_i(s) = \beta_0(s) + \sum_{k=1}^K \xi_{ik}\widehat{\phi}_k(s) \\
        \xi_{ik} \stackrel{\text{iid}}{\sim} N(0, \sigma_k^2)\;, &\hspace{0.25cm}  \text{Cov}(\xi_{ik}, \xi_{il}) = 0 \hspace{0.25cm} \text{for $k \neq l$} 
    \end{align*}
}
\end{algorithm}
\normalsize

\section{Comparison to Weishampel in NHANES}\label{sec:supp_Weishampel_NHANES}

\begin{figure}[!ht]
    \centering
    \includegraphics[width=\textwidth]{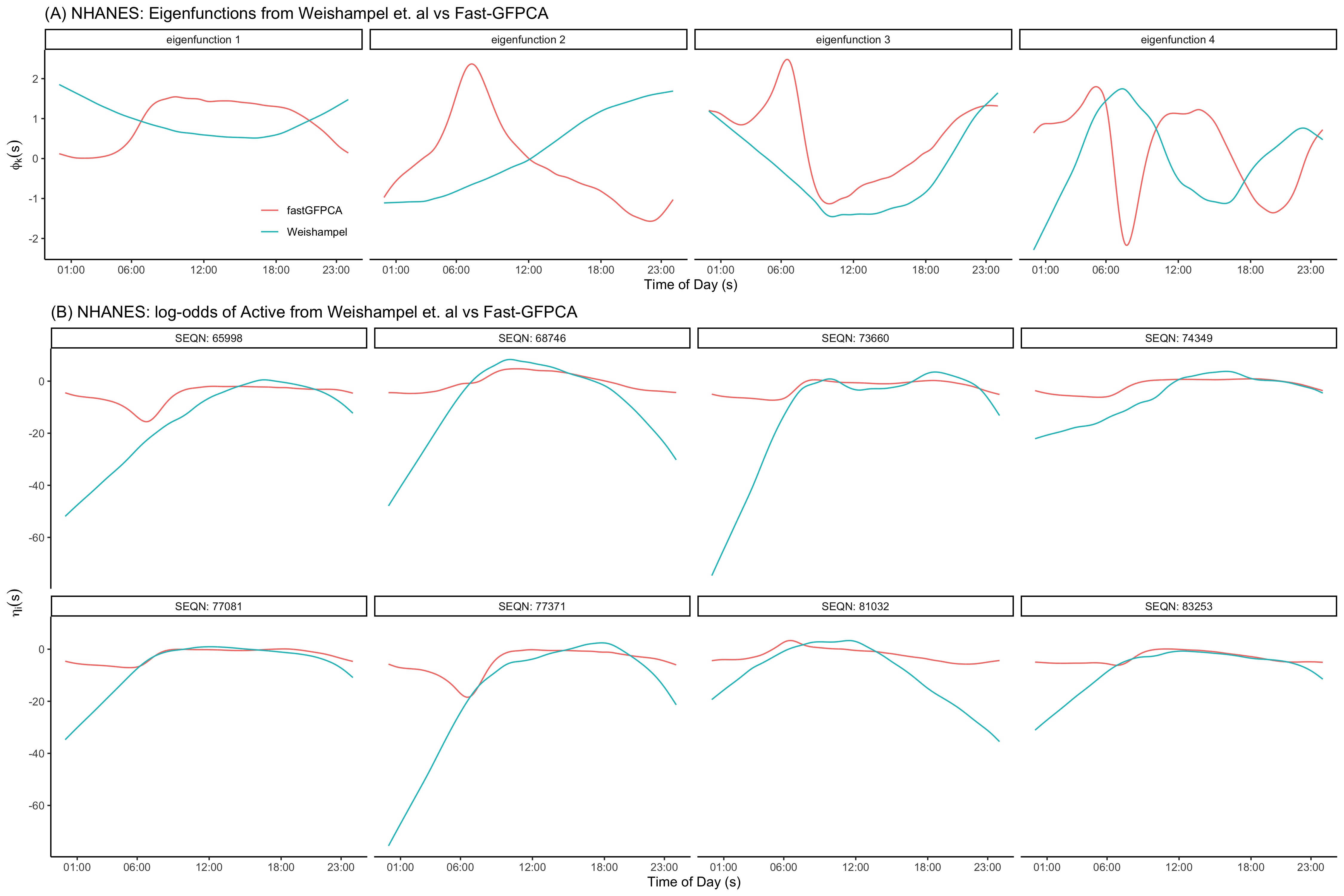}
    \caption{(A) The first four estimated eigenfunctions ($\hat{\phi}_k(s)$) from fast-GFPCA (red lines) versus those obtained from Weishampel et al (blue lines). (B) Estimated log-odds of being active Model estimates are presented as red lines ({\ttfamily fastGFPCA}) or blue lines ({\ttfamily fastGFPCA}). {\ttfamily fastGFPCA} estimates compared with two {\ttfamily fastGFPCA} models. The {\ttfamily fastGFPCA} models (blue columns) are estimated with overlapping windows with bin sizes $w_l =6$ and $w_l =10$.
    \label{fig:weishampel_NHANES}
    }
\end{figure}

Figure~\ref{fig:weishampel_NHANES} presents the results of the GFPCA approach of \cite{WEISHAMPEL2023107647} versus fast-GFPCA applied to the NHANES data. We implemented the \cite{WEISHAMPEL2023107647} approach by first estimating individual models using $30$ penalized cubic regression splines with REML smoothing parameter selection. FPCA on the latent scale was implemented using {\ttfamily refund::fpca.face()}. There are substantial differences in the shapes of these eigenfunctions (Figure~\ref{fig:weishampel_NHANES}(A)), with the first eigenfunction of \cite{WEISHAMPEL2023107647} (Figure~\ref{fig:weishampel_NHANES}(A), left panel, blue line)being dominated by the behavior at the tails of the functions (early AM and late PM). This is possibly due to the fact that many individuals' have all $0$'s (inactive) during the early AM (and to a lesser extent the late PM). When fitting individual GAMs to these data to estimate participants' latent functions, this tends to result in estimates which diverge at the boundary. This effect can be seen in the estimated log odds of being active obtained from \cite{WEISHAMPEL2023107647} in Figure~\ref{fig:weishampel_NHANES}(B). The approach of \cite{WEISHAMPEL2023107647} is compared to the results from fast-GFPCA for $8$ randomly selected participants (each panel), with color indicating estimation approach (blue for Weishampel \cite{WEISHAMPEL2023107647}, red for fast-GFPCA). Each of the estimates from the approach of \cite{WEISHAMPEL2023107647} diverges at the boundaries, though to varying extents.

\section{Further speeding up fast-GFPCA} 
\label{sec_supp:speed_up}
The primary potential computational bottleneck for Fast-GFPCA is the need to re-estimate the subject-specific scores in Step 4 of the Fast GFPCA algorithm. We have identified three potential avenues for speeding up the current algorithm, the first of which we apply in our data application, but leave detailed investigation of each to future work. The first approach is to estimate Steps 1-3 on the entire population, obtaining population level eigenfunctions for estimating subject-specific random effects, but estimating Step 4 on randomly selected sub-samples of the data. In Step 4 all that is being done is re-estimating the population mean function and projection of the subject specific latent estimates onto the population basis estimates. For large sample sizes, such as in our data application, both of these quantities should be relatively stable in sufficiently large sub-samples (e.g. $N=1000$). 

The remaining two approaches are based on the observation in our simulations and data application that the scores obtained in Step 3 are almost perfectly correlated with those from Step 4, but are off by linear scaling factor (See Figure~\ref{fig:need_for_step_4}(B)). This scaling factor appears to be the same or similarly valued across each of the eigenfunctions  Note the slope is nearly identical across eigenfunctions in Figure~\ref{fig:need_for_step_4}(B)). This suggests that if one were able to estimate the scaling factor, final estimates for the subject specific scores would require less computation. We see this as achievable in at least two ways. First, one may estimate Step 4 on a subset of the data, then apply the scaling factor directly to the estimates obtained form Step 3. Alternatively, one may apply the previously suggested subset analysis to estimate the variance of subject scores, $\lambda_k$, and then treat both the estimated eigenfunctions ($[\{\hat{\phi}_k(s)\}, 1 \leq s \leq J]$) and the variance parameters as fixed. In this way, the model in Step 4 requires estimating only the population mean ($\beta_0(s)$), substantially reducing the computational complexity of Step 4. This can be done easily in the {\it mgcv} package, and we illustrate this idea in our data application, though we leave rigorous examination of this idea for future work.

\section{Additional Simulation Results}
\label{sec_supp:additional_sim}

\begin{figure}[!ht]
    \centering
    \includegraphics[scale=0.25]{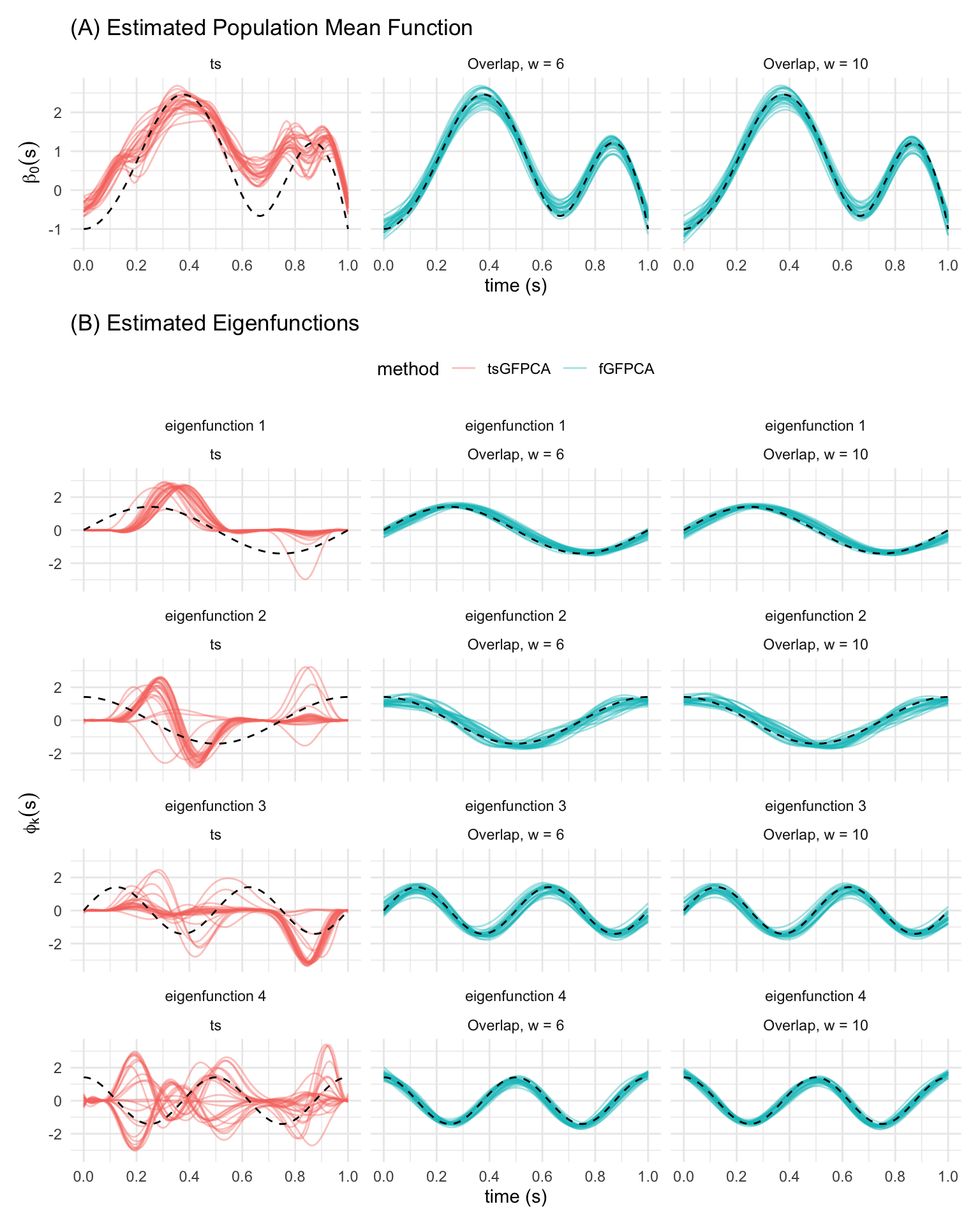}
    \caption{The estimated population mean function ($\hat{\beta}_0(s)$) and the first four estimated eigenfunctions ($\hat{\phi}_k(s)$) from 25 simulated Poisson functional datasets with $N = 100$ subjects and $J = 100$ time points. Model estimates are presented as red lines ({\ttfamily fastGFPCA}) or blue lines ({\ttfamily fastGFPCA}). {\ttfamily fastGFPCA} estimates compared with two {\ttfamily fastGFPCA} models. The {\ttfamily fastGFPCA} models (blue columns) are estimated with overlapping windows with bin sizes $w_l =6$ and $w_l =10$.
    \label{fig:simulation_accuracy_poisson}
    }
\end{figure}

\subsection{Data Driven Simulation}
\label{sec_supp:data_driven_sim}

To investigate the differences in estimating the mean and principal components between vbFPCA and fast-GFPCA approaches we conducted a short simulation study. First the parameters of model~\eqref{eq:full_conditional} were estimated for $K=4$ eigenfunctions using the non-overlapping fast-GFPCA model using a window size of $w=6$. We generated five data sets with $N=4000$ participants, each observed on the same grid as the original data ($J=1440$). Since we did not find parameter estimates differed in the fast-GFPCA approach by bin width and overlapping versus non-overlapping windows, we only estimate the fast-GFPCA model with $w=6$ and non-overlapping windows. 

\begin{figure}[!ht]
    \centering
    \includegraphics[width=\textwidth]{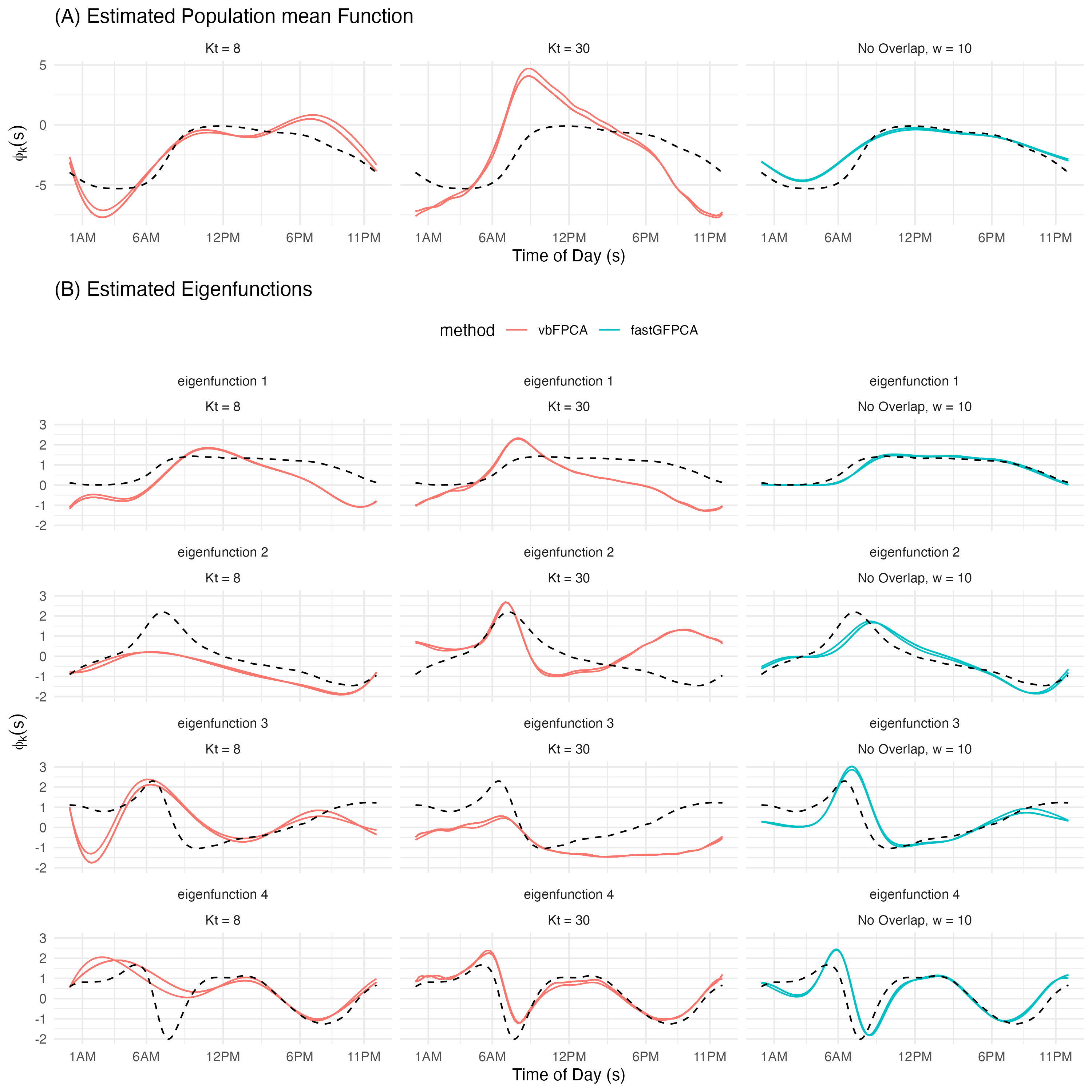}
    \caption{
    Figure showing the estimated population mean function ($\hat{\mu}(s)$) and the first four estimated eigenfunctions ($\hat{\phi}_k(s)$) from 5 simulated datasets based on the NHANES fast-GFPCA results. Model estimates are presented as red lines (vbFPCA) or blue lines (fast-GFPCA). vbFPCA estimates from models using either Kt=8 (left column) or Kt=30 (middle column) basis functions are presented and compared with fast-GFPCA estimated using non-overlapping windows of size $w_l =10$. 
    \label{fig:NHANES_simulation_results}
    }
\end{figure}

Figure~\ref{fig:NHANES_simulation_results} plots the results of this small simulation study and is structured the same as Figure~\ref{fig:NHANES_results_estimated_parameters}, with the estimated population mean function $\hat{\mu}(s)$ and eigenfunctions $\hat{\phi}_k(s)$ plotted in the first and second through fifth rows, respectively. The true mean and eigenfunctions are plotted as solid black lines, while the estimated values across the 5 simulated datasets are plotted as semi-transparent red (vbFPCA) and blue (fast-GFPCA) models, respectively. We see that both the population mean function and eigenfunction estimates are biased in the vbFPCA approach and the bias aligns remarkably well with the estimates we obtained from the real data. In contrast, fast-GFPCA estimates show minimal bias. As expected with the relatively large sample size $N=4000$), estimates of both the population mean and eigenfunctions are extremely consistent across simulated datasets. These findings suggest that the vbFPCA approach may indeed be providing biased parameter estimates in our data application.

\subsection{Simulation results for non-periodic functional data}

While our paper focuses mainly on the simulation results for data simulated with periodic endpoints, we also implemented simulation scenarios with non-periodic curves. The results of these scenarios are summarized in Table \ref{table:supp_case2} below.

\begin{table}[!ht]
    \centering
    \scalebox{0.85}{
    \begin{tabular}{lllllllll}\toprule
\multicolumn{9}{c}{Integrated Squared Error of Model Parameters for Non-Periodic Curves}  \\\cmidrule(lr){1-9} 
                          &                    &                    & Fast-GFPCA     &                &               & vbFPCA        &               & tsGFPCA        \\\cmidrule(lr){4-6} \cmidrule(lr){7-8}\cmidrule(lr){9-9}
\multirow{2}{*}{Estimate} & \multirow{2}{*}{N} & \multirow{2}{*}{J} & \multicolumn{3}{l}{Non-Overlapping Bins}        & K = 8         & K = 30        & K = 30         \\
                          &                    &                    & w = 6          & w = 10         & w = 50        &               &               &                \\\cmidrule(lr){1-9} 
$\psi_(s)$                       & 100                & 100                & 0.95           & 0.98           & -             & 0.54          & 1.28          & \textbf{0.53}  \\
                          &                    & 500                & 0.55           & 0.49           & 1.41          & \textbf{0.31} & 0.52          & -              \\
                          &                    & 2000               & 0.63           & 0.5            & \textbf{0.4}  & 0.52          & 6.85          & -              \\
                          & 500                & 100                & 0.44           & 0.48           & -             & \textbf{0.16} & 0.28          & 0.22           \\
                          &                    & 500                & 0.24           & 0.18           & 1.08          & \textbf{0.09} & 0.14          & -              \\
                          &                    & 2000               & 0.21           & \textbf{0.14}  & \textbf{0.14} & 0.36          & 6.08          & -              \\
                          & 1000               & 100                & 0.34           & 0.42           & -             & \textbf{0.1}  & 0.17          & -              \\
                          &                    & 500                & 0.14           & 0.12           & 0.97          & \textbf{0.05} & 0.09          & -              \\
                          &                    & 2000               & 0.1            & \textbf{0.06}  & 0.08          & 0.35          & 5.92          & -              \\\cmidrule(lr){1-9} 
$\beta_0(s)$                      & 100                & 100                & 11.82          & 13.57          & -             & 16.29         & 28.51         & \textbf{10.09} \\
                          &                    & 500                & \textbf{12.94} & 13.07          & 12.55         & 13.16         & 15.71         & -              \\
                          &                    & 2000               & 11.88          & \textbf{10.74} & 13.2          & 12.84         & 13.08         & -              \\
                          & 500                & 100                & 1.68           & \textbf{1.66}  & -             & 3.71          & 6             & 2.24           \\
                          &                    & 500                & 1.69           & 1.63           & \textbf{1.59} & 3.29          & 3.87          & -              \\
                          &                    & 2000               & 1.96           & 1.93           & \textbf{1.87} & 3.33          & 3.22          & -              \\
                          & 1000               & 100                & \textbf{0.73}  & 0.78           & -             & 1.89          & 3.09          & -              \\
                          &                    & 500                & 0.77           & 0.76           & \textbf{0.71} & 1.5           & 1.77          & -              \\
                          &                    & 2000               & 0.83           & 0.74           & \textbf{0.7}  & 2.21          & 1.76          & -              \\\cmidrule(lr){1-9} 
$\eta_i(s)$                 & 100                & 100                & \textbf{2.31}  & 2.32           & -             & 2.44          & 3.14          & 2.4            \\
                          &                    & 500                & \textbf{0.56}  & \textbf{0.56}  & 0.58          & 0.58          & 0.73          & -              \\
                          &                    & 2000               & 0.16           & 0.16           & 0.17          & \textbf{0.15} & 0.19          & -              \\
                          & 500                & 100                & 2.29           & \textbf{2.3}   & -             & \textbf{2.3}  & 2.42          & 2.32           \\
                          &                    & 500                & 0.59           & 0.59           & 0.61          & \textbf{0.55} & 0.58          & -              \\
                          &                    & 2000               & 0.16           & 0.16           & 0.17          & \textbf{0.15} & \textbf{0.15} & -              \\
                          & 1000               & 100                & 2.3            & 2.31           & -             & \textbf{2.29} & 2.34          & -              \\
                          &                    & 500                & 0.57           & 0.58           & 0.6           & \textbf{0.55} & 0.56          & -              \\
                          &                    & 2000               & \textbf{0.15}  & \textbf{0.15}  & 0.16          & \textbf{0.15} & \textbf{0.15} & -   \\          
\bottomrule\end{tabular}
    }
    \caption{
    Mean integrated squared error (MISE) for $\psi_(s)$ and $\hat{\eta}_i(s)$ and integrated squared error (ISE) for $\beta_0(s)$ across methods and simulation scenarios. In each row the method(s) with the lowest MISE for that simulation scenario is in \textbf{bold}. All data summarized in this table were simulated with population mean $\beta_0(s) = 0$ and with non-periodic eigenfunctions. An ``-'' indicates that that model was not evaluated for a given simulation scenario.
    \label{table:supp_case2}
    }
\end{table}

\end{document}